\documentclass[10pt,a4paper,twocolumn]{article}
\usepackage[utf8x]{inputenc} 
\usepackage{amsmath}
\usepackage{amsfonts}
\usepackage{amssymb}
\usepackage{upgreek}
\usepackage[left=2cm,right=2cm,top=2cm,bottom=2cm]{geometry}
\usepackage[T1]{fontenc} % Use modern font encodings
\usepackage{color}
\usepackage{graphicx} 
\usepackage{caption}
\usepackage{epstopdf}
\usepackage{hyperref}
\usepackage{ulem}
\usepackage{multicol}
\usepackage[numbers,sort&compress]{natbib}
\usepackage{abstract}
\usepackage[font=small,labelfont=bf]{caption}
\usepackage{multirow}
\usepackage[resetlabels]{multibib}
\newcites{SI}{References}
\usepackage{etoc}
\newcommand{\specificthanks}[1]{\@fnsymbol{#1}}

\title{Electrochemistry, Ion Adsorption and Dynamics in the Double Layer: A Study of NaCl(aq) on Graphite}

\author{Aaron R. Finney,\thanks{To whom correspondence may be addressed: a.finney@ucl.ac.uk; m.salvalaglio@ucl.ac.uk}~$^,$\thanks{Thomas Young Centre and Department of Chemical Engineering, University College London, London WC1E 7JE, United Kingdom} \space Ian J. McPherson,\thanks{Department of Chemistry, University of Warwick, Coventry, CV4 7AL, United Kingdom} \space  Patrick R. Unwin$^{\ddagger}$  and Matteo Salvalaglio$^{\ast,\dagger}$}

\date{}

\begin{document}
\etocdepthtag.toc{mtchapter}
\etocsettagdepth{mtchapter}{subsection}
\etocsettagdepth{mtappendix}{none}
\twocolumn[
\maketitle

\begin{onecolabstract}
\noindent
Graphite is a ubiquitous electrode material with particular promise for use in e.g., energy storage and desalination devices, but very little is known about the properties of the graphite--electrolyte double layer at technologically relevant concentrations.  
Here, the (electrified) graphite--NaCl(aq) interface was examined using constant chemical potential molecular dynamics (CμMD) simulations; this approach avoids ion depletion (due to surface adsorption) and maintains a constant concentration, electroneutral bulk solution beyond the surface.
Specific Na$^+$ adsorption at the graphite basal surface causes charging of the interface in the absence of an applied potential.
At moderate bulk concentrations, this leads to accumulation of counter-ions in a diffuse layer to balance the effective surface charge, consistent with established models of the electrical double layer.
Beyond ∼0.6 M, however, a combination of over-screening and ion crowding in the double layer results in alternating compact layers of charge density perpendicular to the interface. 
The transition to this regime is marked by an increasing double layer size and anomalous negative shifts to the potential of zero charge with incremental changes to the bulk concentration.
Our observations are supported by changes to the position of the differential capacitance minimum measured by electrochemical impedance spectroscopy, and are explained in terms of the screening behaviour and asymmetric ion adsorption.
Furthermore, a striking level of agreement between the differential capacitance from solution evaluated in simulations and measured in experiments allows us to critically assess the accepted norm that electrochemical capacitance measurements report simply on the density of states of the graphite material.
Instead, our work shows that the solution side of the double layer provides the more dominant contribution (hitherto neglected).
Finally, ion crowding at the highest concentrations (beyond ∼5 M) leads to the formation of liquid-like NaCl clusters confined to highly non--ideal regions of the double layer, where ion diffusion is up to five times slower than in the bulk. 
The implications of changes to the speciation of ions on reactive events in the double layer are discussed.
\end{onecolabstract}
]
\saythanks

\section{Introduction}
 Carbon--electrolyte interfaces are fundamental to the operation of many technological devices, particularly in the areas of energy storage (e.g., supercapacitors), filtration and sensing. \cite{simon_materials_2008,choi_challenges_2012,cohen-tanugi_water_2012,aghigh_recent_2015,cong_graphene-based_2014,martin-yerga_correlative_2019} 
 More generally, understanding chemical activity in the vicinity of  interfaces is important for catalysis, corrosion and crystallisation. \cite{magnussen_toward_2019,gros_modelling_2019,alkire_electrochemistry_2015,landolt_corrosion_2007,kashchiev_nucleation_2000} 
 At the interface, variations in interatomic forces induce perturbations to both the structure and dynamics of solute and solvent molecules over a finite volume of the solution known as the \textit{double layer}. 
 When the solute is ionised, and particularly when the surface is charged, any net accumulation of one particular ion in the double layer leads to a departure from electroneutrality and to the generation of local electric fields.

% The double layer plays a significant role in electrochemically driven processes at surfaces. \cite{magnussen_toward_2019} Indeed, understanding how solvent and electrolytes in the double layer control surface activities (at a range of electrolyte concentrations) is key to ensuring the effective design of filters and catalysts.\cite{cohen-tanugi_water_2012,bohra_modeling_2019} The influence of the double layer extends further, however. For example, interfaces are already known to increase the rates of crystal nucleation, compared to those in the bulk, by orders of magnitude. \cite{vekilov_nucleation_2010,sosso_crystal_2016} Classical nucleation theory attributes this control to the adsorption of crystalline embryos onto solid substrates which reduces the unfavourable surface free energy for the emerging phase. \cite{kashchiev_nucleation_2000} In this context, variations in the supply of monomers within the double layer and their activities in this region should also be considered.

Many details about the carbon-aqueous electrolyte interface remain unresolved. 
Most experimental studies of these systems focus on simple aqueous binary electrolyte solutions in contact with highly ordered pyrolytic graphite (HOPG): a synthetic graphite which exposes relatively large, atomically flat sp$^2$ hybridised carbon planes. \cite{alkire_electrochemistry_2015} 
Experimental data on the HOPG-aqueous electrolyte interface structure is mainly limited to differential capacitance ($C^d$) measurements,\cite{randin_differential_1972,zou_investigation_2016,iamprasertkun_capacitance_2019}
While the minimum value of $C^d$ was attributed to the dominant role of the electronic properties of graphite, increasing $C^d$ beyond this region of potential is associated to the structure of the solution. \cite{randin_differential_1971,gerischer_density_1987,luque_electric_2012}
The Gouy-Chapman-Stern (GCS) model---see SI Section \ref{section:theory} for more details---predicts a compact layer of counter-ions adjacent to an electrified, planar surface, followed by a diffuse solution layer enriched in counter-ions and depleted in co-ions.\cite{gouy_sur_1910,chapman_li_1913,stern_zur_1924} 
At low electrolyte concentrations and low applied surface potentials, this model effectively predicts $C^d$; \cite{torrie_electrical_1980,schmickler_new_1986} even if the simplified interfacial geometry it implies is unphysical for the electrode/electrolyte interface.\cite{schmickler_double_2020} 
However, at higher concentrations---relevant to many technical applications--- deviations from the model are observed. 
For example, at 0.5~M and beyond, $C^d$ depends upon the cation type, and asymmetries in the $C^d$--potential curves are observed in experiments studying alkali chloride solutions in contact with HOPG. \cite{iamprasertkun_capacitance_2019,iamprasertkun_understanding_2020}

Ion-specific changes to $C^d$ were previously associated to differences in the charge separation distances between cations in the compact layer and HOPG. \cite{iamprasertkun_capacitance_2019} 
A more complex interpretation, considering cation (de)solvation energies and charge transfer with the carbon surface, was also 
proposed. \cite{zhan_specific_2019} 
A $C^d$ dependence due to the interaction of cations with the electrode seemingly contrasts with other experimental observations, however. 
For example, a greater capacitance at positively charged HOPG surfaces in LiCl(aq) was ascribed to increased densities of chloride ions in the double layer. \cite{zou_investigation_2016}
Furthermore, concentration-dependent shifts in the potential associated with the minimum in $C^d$ is considered indicative of specific adsorption, \cite{bard_electrochemical_2001} and in the case of KF(aq),\cite{iamprasertkun_understanding_2020} and NaCl(aq) (described herein), this shift is negative, consistent with the increasing accumulation of anions in the compact layer. 
The conflict in these interpretations exists only in the context of the GCS model, however, and should be reconciled in models that allow greater complexity in the structure of the double layer to emerge at high electrolyte concentrations.

Modifications to the Poisson-Boltzmann-based models of the double layer, such as GCS, were proposed; \cite{borukhov_steric_1997,fedorov_ionic_2008,uematsu_effects_2018,howard_behavior_2010} these take into account the aforementioned ion correlations, their specific adsorption and steric effects. 
Nonetheless, the complexity of the system favours the application of atomistic simulations to fully characterise the double layer structure that results from changes to bulk electrolyte concentrations and applied surface potentials.
Calculations at the level of density functional theory with implicit solvents have provided significant insight into the nature of charge screening in the double layer. \cite{zhan_computational_2017,zhan_specific_2019}. 
Molecular dynamics (MD) simulations adopting classical force fields, however, are the preferred tool to investigate the structure \textit{and} dynamics of electrolytes in explicit solvents in contact with graphitic surfaces. \cite{cole_ion_2011,jiang_molecular_2016,williams_effective_2017,shi_ion_2013,chen_effect_2014,bo_molecular_2015,yang_edge_2016,zhan_enhancing_2016}
Based on density functional theory (DFT) calculations, classical pairwise interaction potentials were parameterised;\cite{williams_effective_2017,shi_ion_2013,chen_effect_2014} these capture the polarisability of the solution and carbon at the interface, which can  play a significant role in structuring the double layer. \cite{gschwend_discrete_2020,gschwend_how_2020}
Such classes of models indicate an asymmetric adsorption of ions; the implications of this on the electrochemical properties of the double layer will be addressed in the present article by combining our own simulations and experimental measurements.

Constant chemical potential MD (C$\mu$MD) \cite{perego_molecular_2015} simulations were performed to provide much needed insight into the response of the structure and dynamics of ions in the double layer of the (electrified) graphite--NaCl(aq) system over a wide range of bulk electrolyte concentrations.
C$\mu$MD mimics open boundary conditions; thus, maintaining a constant thermodynamic driving force for ion adsorption at graphite, and conserving electroneutral solutions beyond the double layer. 
With this approach, we are able to relate the spatial extent of the double layer to the nature of charge screening in this region. 
This screening also determines changes to the electrochemical properties of the double layer.
In addition, we obtain a detailed description of the local (electro)chemical potential, speciation and mobility of ions orthogonal to the surface. 
The results indicate a significant departure from \textit{ideal} solution behaviour in regions confined to the double layer even at moderate levels of NaCl(aq) concentration in the bulk.

\section{Results and Discussion}
\subsection{The Structure of NaCl(aq) Solutions at the Graphite Surface}

\begin{figure*}[tb]
\centering
  \includegraphics[width=0.5\linewidth]{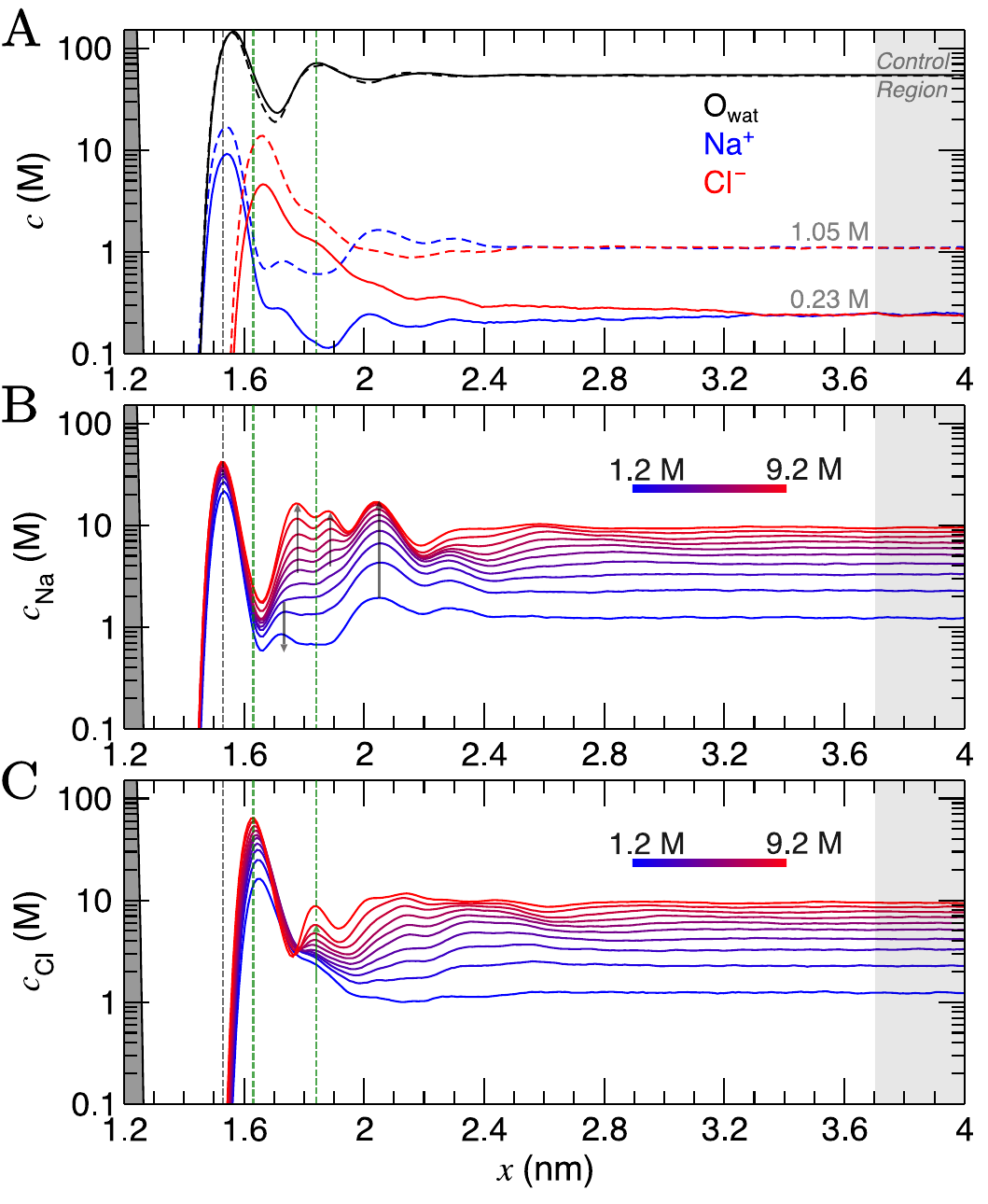}
  \caption[Ion and water concentration profiles]{Ion and water molar concentration ($c$) profiles from C$\mu$MD simulations as a function of $x$: the distance from the centre of the simulation cell. The graphite carbon is shown by the grey peak on the left of the $x$ axis. A provides Na$^+$ (blue), Cl$^-$ (red) and water oxygen ($\mathrm{O_{wat}}$; black) concentrations when the concentration of ions in the bulk is 0.23 M (solid line) and 1.1 M (dashed line). B and C provide the cation (top panel) and anion (bottom panel) concentrations as a function of $x$ in simulations targeting the higher end of the total concentration range. The colour scale here indicates the mean molarity of ions in the bulk. The shaded dashed lines mark the maximum in the first sodium peak (grey) and first two chloride peaks (green) at the highest bulk concentration. Arrows are provided to highlight the changes in the profiles as ion concentrations in the bulk are increased. A 0.3~nm excluded region separates the edge of the graphite basal plane from the ionic solution; this is due to atom centres being used to calculate the concentration profiles.}   
  \label{fgr:density-profiles}
\end{figure*}

Simulation cells were prepared where a graphite slab, comprising eight graphene layers, was positioned at the centre of the simulation cell $x$ axis in contact with NaCl(aq), such that the system was symmetrical about $x=0$. 
All methodological details are provided in SI Section \ref{sec:methods}. 
C$\mu$MD simulations were performed using the GROMACS (v 2018.6) MD package \cite{hess_gromacs_2008} with the Plumed (v 2.5) Plugin \cite{tribello_plumed_2014} for a range of fixed bulk solution concentrations: $c^{\mathrm{b}}_{\mathrm{NaCl}}=$0.23--1.05~M and 1.2--9.2~M in two system set-ups. 
The upper limit here significantly exceeds the solubility of NaCl in water, and would not be possible to prepare experimentally, but is instructive to study computationally. 
The nanosecond timescales associated with MD give rise to vanishingly low probabilities for crystal nucleation and allow the metastable solution state to be investigated. 
Within minimal fluctuations, the C$\mu$MD method successfully maintained the bulk concentration of cations and anions beyond the interface (see SI Section \ref{sec:concentration}).

\paragraph{Concentration profiles orthogonal to the surface.} 
Concentrations as a function of $x$ for species in the solution phase are provided in Figure \ref{fgr:density-profiles} and also in full in Figures \ref{fgr:rho-profiles1} and \ref{fgr:rho-profiles2}. 
Preferential adsorption of Na$^+$ was observed at the graphite surface over the entire concentration range sampled, in line with the predicted order of ion adsorption energies using the adopted force field.\cite{williams_effective_2017} 
Solvated Na$^+$ ions directly coordinate to graphite, shown by the narrow peak at $x=1.5$~nm in Figure \ref{fgr:density-profiles}~A. 
Concentrations exceed $c_\mathrm{Na}^\mathrm{b}$ by approximately two orders of magnitude even at the lowest $c_\mathrm{NaCl}^\mathrm{b}$ (0.23 M). 
An adjacent layer of Cl$^-$ ions was observed, separated by 0.1~nm in terms of maximum concentrations, and in line with other simulations \cite{chen_multiscale_2018}.
The Cl$^-$ concentration in this layer at 0.23~M also significantly exceeds the $c_\mathrm{Cl}^\mathrm{b}$, although its broader width highlights a diffuse ordering of the anion. 
Considering the adsorbed layer of Na$^+$ to represent an effective surface charge density, this picture of charge screening is qualitatively consistent with the predictions of the GCS model; with the Cl$^-$ diffuse region of the double layer representing the counter-ion charge in solution. 
However, no clear boundary between these two regions is apparent in the concentration profiles, and any binary assignment of surface-bound states neglects the complexity of the dynamic adsorption in the first ion layers.

As the bulk concentration of ions is raised (see Figure \ref{fgr:rho-profiles1}), a clear departure from the above screening behaviour is observed. 
Around $c^{\mathrm{b}}_{\mathrm{NaCl}}$ = 0.5~M, the Cl$^-$ peak narrows and a second, diffuse layer of cations around $x=2$~nm emerges that exceeds bulk ion concentrations. 
Figure \ref{fgr:density-profiles}~A highlights that at $c^{\mathrm{b}}_{\mathrm{NaCl}}=1.05$~M, the concentration of cations exceeds that of anions at $x=2$--2.4~nm, and a hierarchical ordering of ions with opposing charge is apparent. 
A more compact double layer region is evident at the highest concentrations (see Figures \ref{fgr:density-profiles}~B and C, with the complete data set available in Figure \ref{fgr:rho-profiles2}).
In the range $c^{\mathrm{b}}_{\mathrm{NaCl}}=1$--9~M, a shift in the position of the Cl$^-$ first peak by $\Delta x\approx -0.03$~nm highlights a contraction of the first ion layers, and the diffuse Cl$^-$ peak at 1.2~M was resolved into two clear peaks (with a second peak emerging at $x=1.83$~nm at around 6~M bulk ion concentrations).
This is concomitant with a shifting of the second Na$^+$ peak away from the graphite surface, which ultimately splits into a rather diffuse doublet peak which confines the Cl$^-$ layer. 

%%The crowding of ions in the double layer region at graphite was observed in other simulation studies. \cite{chen_multiscale_2018,elliott_qmmd_2020,yang_edge_2016,cole_ion_2011} 
The crowded structure at higher concentrations is reminiscent of the double layer in molten LiCl at planar electrodes under the constraint of a constant applied potential. \cite{vatamanu_molecular_2010} 
Indeed, in double-layer capacitors containing ionic liquids, steric crowding at the electrode is a common feature, \cite{kornyshev_double-layer_2007,vatamanu_charge_2017} which has been confirmed by atomic force microscopy (AFM). \cite{hayes_double_2011,zhang_structural_2020} 
The similar response of the double layer structure to changes in bulk concentrations and applied surface potentials was identified in simulations over 40 years ago. \cite{torrie_electrical_1980} 
As well as the increasingly non-monotonic concentration profiles (shown clearly in Figure \ref{fgr:max-concn}~A on increasing $c^{\mathrm{b}}_{\mathrm{NaCl}}$), 
a shift in the ratio of maximum Na$^+$:Cl$^-$ concentrations occurs around 3~M due to the narrowing of the second Cl$^-$ peak (see Figure \ref{fgr:max-concn}~B) which represents a significant departure from the double layer structure at the lowest concentrations.

Perturbations to the solvent structure were apparent, mainly due to the presence of graphite. 
Three peaks are observed (see Figure \ref{fgr:density-profiles}~A) in the water concentration profiles at the lower end of the bulk ion concentration range simulated. Solvent layers are separated by 0.3~nm; this distance was determined for water at clean graphite surfaces in a recent study combining simulations with AFM measurements. \cite{uhlig_atomically_2021}
Only limited ordering of the orientation of water molecules perpendicular to the graphite surface was observed. 
This is highlighted in Figure \ref{fgr:rho-profiles1}, which shows that the peaks for water O and H atoms are approximately at the same position in $x$. 
Maximum values of $c_\mathrm{Cl}$ at the interface tended to be found where the densities of water oxygen atoms are close to a minimum. 
Additional ion layers at the highest concentrations induce additional complexity to the water structure in the double layer (see Figure \ref{fgr:rho-profiles2}). 

\begin{figure*}[tb]
\centering
  \includegraphics[width=0.55\linewidth]{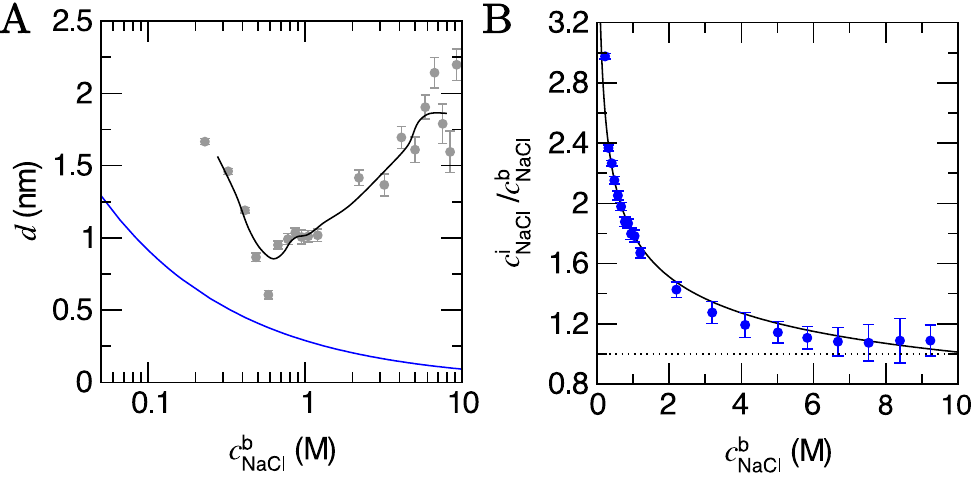}
   \caption[Double Layer Size]{A: The depth of the double layer region, $d$, in C$\mu$MD simulations with varying bulk ion concentrations, $c^\mathrm{b}$. The grey points provide the size of the double layer in $x$, with error bars indicating uncertainties of one standard deviation in the data. The black line provides a moving average for the data. Note that the minimum value of $x$ for the solution phase (with reference to Figure \ref{fgr:density-profiles}) was taken to be the mid-point between the carbon surface and the minimum value in ion concentration profiles: $x \approx 1.35$ nm. Also shown in the figure is the blue curve which is the Debye length (see SI Section \ref{section:theory}). B provides the normalised, mean interface ion concentrations ($c^{\mathrm{i}}_{\mathrm{NaCl}}$) in a 1.5~nm solution region close to the graphite surface vs. the bulk concentration of ions. The black line is a fit to the data with functional form $1.8(c^{\mathrm{b}}_{\mathrm{NaCl}})^{-0.25}$ and error bars show uncertainties of one standard deviation.}
  \label{fgr:edl-depth}
\end{figure*} 

\paragraph{Extent of the double layer.} 
The edge of the double layer region on approach to the surface was marked by the position where solutions deviate from electroneutrality; hence, $\langle{c_{\mathrm{Na}}(x)}\rangle \neq \langle{c_{\mathrm{Cl}}(x)}\rangle$ (where angular brackets indicate the mean concentrations in 0.5~nm moving windows in $x$).
Figure \ref{fgr:edl-depth}~A indicates that the double layer contracts as the ionic strength in the bulk solution initially increases, reaching a minimum around $c^{\mathrm{b}}_{\mathrm{NaCl}}=0.6$~M. 
Beyond this bulk concentration, the double layer size increases, plateauing around 2~nm above 5~M, with some noise in the data.
Overall, using this composition measure, the double layer size is $0.6$--$2.2$~nm. 

The Debye length ($\kappa^{-1}$) is the characteristic length over which the electrostatic effect of a charge carrier in solution decays. 
This is derived from a linearised Poisson-Boltzmann equation (see Section \ref{section:theory} for details), and is assumed to accurately determine the size of the double layer at low electrolyte concentrations.
Figure \ref{fgr:edl-depth} indicates that $\kappa^{-1}$ decreases monotonically as $c^{\mathrm{b}}_{\mathrm{NaCl}}$ increases, and beyond 0.084~M, $\kappa^{-1}$ is below 1~nm. 
At relatively low concentrations, one can reconcile the decreasing double layer size by considering that increasing charge densities close to the surface will lead to a less diffuse double layer as contraction of the layers occurs. 
At high concentrations, however, the ionic crowding near the surface induces further perturbations to the solution away from the interface, and the double layer size increases with $c^{\mathrm{b}}_{\mathrm{NaCl}}$.
A theoretical framework to predict the `capacitive compactness' of the double layer was recently presented; \cite{guerrero-garcia_quantifying_2018} this indicates the dependency of the size of the interfacial region upon the ion valency.  

It is instructive to consider the size of the interface region where the solvent structure is perturbed (cf. the bulk), which turns out to be independent of $c^{\mathrm{b}}_{\mathrm{NaCl}}$ over the entire concentration range sampled.  
SI Section \ref{section:dlsize} details these measurements which indicate an interface region that is $1.4 \pm 0.3$~nm in size, that is approximately 3--5 water layers from the graphite surface. 
The structuring of water at planar interfaces appears to be rather insensitive to the electrolyte concentration and, for most practical purposes, the substrate material and surface contamination. \cite{uhlig_atomically_2021}

\paragraph{Mean ion concentrations in the double layer.} 
NaCl concentrations at the interface ($c^{\mathrm{i}}_{\mathrm{NaCl}}$) can be measured by integrating  $(c_{\mathrm{Na}}(x)c_{\mathrm{Cl}}(x))^\frac{1}{2}$ in regions of the profiles in Figure \ref{fgr:density-profiles}. 
To ensure a fair comparison between different cases of bulk concentration, a 1.5~nm solution region closest to the graphite surface was integrated, and the $c^{\mathrm{i}}_{\mathrm{NaCl}}$ normalised by $c^{\mathrm{b}}_{\mathrm{NaCl}}$ are provided in Figure \ref{fgr:edl-depth}~B. 
The plot shows a rapid decay in $c^{\mathrm{i}}_{\mathrm{NaCl}}/c^{\mathrm{b}}_{\mathrm{NaCl}}$ on increasing $c^{\mathrm{b}}_{\mathrm{NaCl}}$, with concentrations at the interface converging to those in the bulk when $c^{\mathrm{b}}_{\mathrm{NaCl}}> \sim6$~M. 

At the lowest $c^{\mathrm{b}}_{\mathrm{NaCl}}$, the concentrations of ions at the interface are three times greater than those in the bulk and the decay in the relative interfacial concentrations is proportional to $(c^{\mathrm{b}}_{\mathrm{NaCl}})^{-0.25}$ (see Figure \ref{fgr:edl-depth}~B). 
While finite ion size effects clearly play a role in the local ion concentrations in the double layer, the total concentrations of ions in this region vary continuously with bulk concentration and can therefore be predicted without the need for simulations at specific concentrations. 
It is important to note that the total concentrations of cations and anions over the entire double layer region are equal, as shown in Figure \ref{fgr:int-concn}, and that significant ordering at the highest bulk concentrations means that, locally, ion concentrations can significantly exceed the bulk (see Figure \ref{fgr:max-concn}~A).

\begin{figure*}[tb]
\centering
  \includegraphics[width=0.75\linewidth]{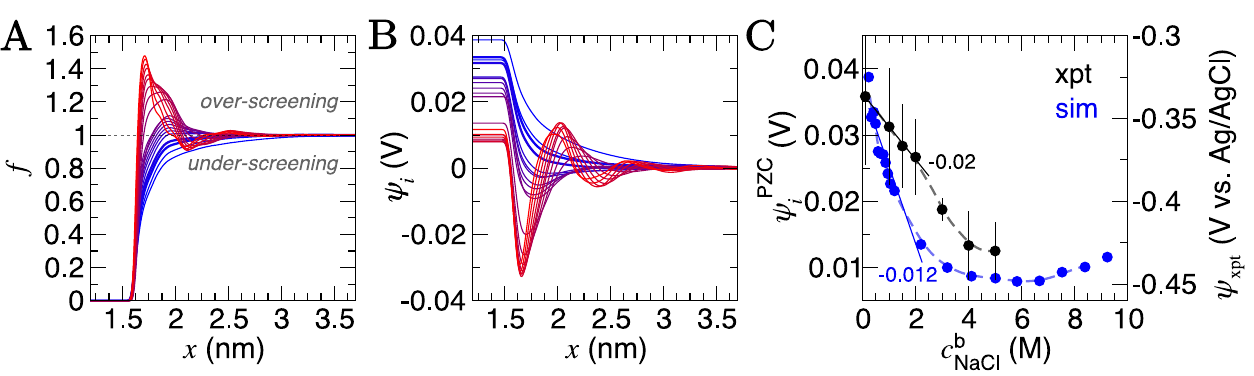}
   \caption[Double Layer Electrostatics]{Charge screening in the double layer. A provides the ion charge screening factor, $f$, as a function of distance, $x$, from the graphite surface. B provides the electric potential, $\psi$, calculated from ion charge distributions in the same region. The colour scale from blue to red in A and B indicates increasing bulk ion concentrations. C provides the potential difference across the interface region as a function of $c_{\mathrm{NaCl}}^{\mathrm{b}}$. The simulation data, taken from the difference in $\psi_i$ in panel B, are shown by the blue circles (the left $y$ axis scale apply to these data only), and measurements from electrochemical experiments are provided by the black circles (the right $y$ axis scale applies here).    Statistical uncertainties of one standard deviation in the data are shown by the error bars. The dashed lines provide a moving average of the data and the solid lines are a linear fit to the data when $c_{\mathrm{NaCl}}^{\mathrm{b}}<3$~M, the gradient magnitudes for which are provided.}
  \label{fgr:pzc-info}
\end{figure*}

\paragraph{Electrical properties of the double layer.} 
Even at uncharged graphite surfaces, asymmetries in the adsorption of ions leads to deviations from local electroneutrality, as shown by fluctuations in the solution charge density, $\rho$, as a function of $x$. 
Following the Poisson equation,
\begin{equation}
    \frac{d^2\psi(x)}{dx^2} = \frac{d\mathrm{E}(x)}{dx} = \frac{-\rho(x)}{\varepsilon }
\label{eqn:poisson-main}
\end{equation}
this leads to varying electric fields, $\mathrm{E}$, and electric potential, $\psi$ orthogonal to the surface. 
In the above equation, $\varepsilon$ is the permittivity of the medium ($\varepsilon=\varepsilon_0\varepsilon_r$ where $\varepsilon_0$ and $\varepsilon_r$ refer to the the permittivity of vacuum and the relative permittivity, respectively). 
Figure \ref{fgr:poisson-plots} provides these quantities over the full concentration range. These indicate that, at the the limit of large $x$, $\mathrm{E}(x)$ and $\psi(x)$ converge to zero, corresponding to the solution bulk.

A screening factor, $f$, can be defined as,
\begin{equation}
f(x')=\frac{\int_0^{x'} n_{\mathrm{Cl}}(x) dx}{\int_0^{x'} n_{\mathrm{Na}}(x) dx}
\label{eq:screeningfactor}
\end{equation}
with $n_{\mathrm{Na}}$ appearing in the denominator as this is the ion that consistently adsorbs in the first solution layer next to the substrate. 
Figure \ref{fgr:pzc-info}~A provides $f(x)$ for the entire concentration range sampled. 
At the lowest $c^{\mathrm{b}}_{\mathrm{NaCl}}$, $f(x)$ increases smoothly and converges to one in the solution bulk, consistent with screening by a diffuse anion layer. 
As $c^{\mathrm{b}}_{\mathrm{NaCl}}$ increases beyond around 0.5~M, and the compensating anion charge layer becomes more compact, \textit{over-screening} of the cation charge occurs and $f(x)>1$. 
Over-screening in molten salts is a phenomenon that has been known for some time. \cite{revere_structure_1986} 
In ionic liquids at electrified interfaces, over-screening was suggested as a possible control on the electrochemical kinetics at the interface. \cite{fedorov_ionic_2014}

The electric potential in $x$ due to the charge distribution of ions ($\psi_i$) can be calculated using Equations \ref{eqn:poisson-main} and \ref{eq:screeningfactor}, noting that $\rho_i(x)=e (n_{\mathrm{Na}}(x)-n_{\mathrm{Cl}}(x))$, where $e$ is the elementary charge and $n$ are ion number densities:
\begin{equation}
\psi_i(x') = \int_0^{x'} \left( - \frac{e}{\varepsilon_0 \varepsilon_r}(1-f(x'))\int_0^{x'} n_{\mathrm{Na}}(x) \; dx \right) dx
\label{eqn:psiscreening}
\end{equation}
The dielectric constant of the medium is affected by the proximity of interfaces \cite{itoh_dielectric_2015,fumagalli_anomalously_2018,bonthuis_dielectric_2011} and ion concentrations \cite{gavish_dependence_2016}. 
In our analyses $\varepsilon_r=71$, the value for SPC/E
water.\cite{fennell_simple_2012} 
A constant dielectric allows us to identify the causal response of the potential difference to changes in the screening factor and bulk electrolyte concentration.
Indeed, when the non-uniform permittivity in $x$ is accounted for, Figure \ref{fgr:poisson-pzc} shows that the potential drop across the double layer is approximately constant (0.4198 $\pm$ 0.004~V) regardless of the bulk concentration of ions, although any trends are very sensitive to fluctuations in the water structure and the numerical precision of partial water atom charges.

The $\psi_i(x)$ curves calculated using Equation \ref{eqn:psiscreening} are provided in Figure \ref{fgr:pzc-info}~B, shifted so that $\psi_i(x)=0$ in the solution bulk. 
When $c^{\mathrm{b}}_{\mathrm{NaCl}}\gtrsim 0.5$~M, the sign of $\psi_i$ alternates due to the crowding of ions in the double layer.
The value of $\psi_i$ at the graphite surface is called the potential of zero charge: $\psi_i^{\mathrm{PZC}}$. 
This potential difference is shown in Figure \ref{fgr:pzc-info}~C as a function of bulk concentration. 
It is clear that the effect of increasing $c^{\mathrm{b}}_{\mathrm{NaCl}}$ is to decrease $\psi_i^{\mathrm{PZC}}$ with a $-0.012$~V/M gradient at moderate bulk concentrations. An inflection point is observed when $c^{\mathrm{b}}_{\mathrm{NaCl}} \approx$~6~M, where further increases in bulk ion concentrations result in positive changes to $\psi_i^{\mathrm{PZC}}$.

The slope, $(d \psi_i^\mathrm{PZC} / d c^{\mathrm{b}}_{\mathrm{NaCl}})_{T,\sigma}$  is related to the so-called Esin-Markov coefficient. \cite{bard_electrochemical_2001} 
This is one of the few conventional means to experimentally assess the extent of specific adsorption, and would seem to provide an ideal way to relate the simulations to experiments. 
Often, $C^d$ data are recorded at widely spaced potentials, limiting the accuracy with which the minimum in $C^d$---often taken to represent $\psi_i^{\mathrm{PZC}}$---is known.\cite{zou_investigation_2016,iamprasertkun_capacitance_2019} 
We therefore measured $C^d$ of freshly exfoliated HOPG as a function of potential over a range of concentrations, with 10 mV potential resolution, and examined the minimum in $C^d$ as a proxy for the PZC (see section \ref{sec:expt}). 
With this fine potential resolution, the $C^d$--$\psi^\mathrm{xpt}$ curve displays two minima within 300 mV; the global minimum becomes deeper and shifts to more negative potentials with increasing concentration (Figure \ref{fgr:mean_Cd}). 
Note that this feature remains visible at 50 mV potential resolution (Figure \ref{fgr:coarse_Cd}), but would be lost at lower resolutions, often reported in the literature.\cite{randin_differential_1972,zou_investigation_2016,iamprasertkun_capacitance_2019} 

Computational models considering the effect of asymmetric ion adsorption indicate a shift to the PZC on incremental changes to $c^{\mathrm{b}}$ that follows the sign of the preferentially adsorbing ion. \cite{vatamanu_molecular_2010}
However, these do not consider the situation of electrolyte solutions in which both cations and anions have favourable, but varying strength of interactions with the surface.
Experimental studies sometimes attribute $C^d$ values solely to the adsorption of ions whose sign is opposite to the sign of the potential change relative to the $C^d$ minimum. \cite{iamprasertkun_capacitance_2019} 
This is a rather simple interpretation under conditions where $c^{\mathrm{b}}$ is far beyond the levels where alternating layers of charge emerge in the double later structure.
The shift in the $C^d$ minimum (see Section \ref{sec:expt} and Figure \ref{fgr:pzc-info}~C) is in good agreement with the shift in PZC found from the simulations, with a moderate negative gradient at lower concentrations.
This contrasts with the linear shift as a function of $\ln{c^{\mathrm{b}}_{\mathrm{NaCl}}}$ seen on Hg electrodes.\cite{grahame_electrical_1947}. 
The agreement provides considerable support to the simulation results.

The implication of Equation \ref{eqn:psiscreening} is that increasing levels of screening, with the same underlying cation density distribution, results in positive changes to $\psi_i^{\mathrm{PZC}}$. 
Conversely, if the screening of charges in the double layer is unchanging and the bulk concentration increases, then $\psi_i^{\mathrm{PZC}}$ becomes more negative. 
In our simulations, Figure \ref{fgr:intscreening} shows that at the limit of large $x$, $\int (1-f(x))$ decreases (due to increased levels of screening in a less diffuse counter-ion charge cloud) and, combined with increasing $c^{\mathrm{b}}_{\mathrm{NaCl}}$, the relatively large negative gradient, $d\psi_i^{\mathrm{PZC}}/dc^{\mathrm{b}}_{\mathrm{NaCl}}$, at the lowest concentrations reduces on increasing $c^{\mathrm{b}}_{\mathrm{NaCl}}$. 
At the highest concentrations, however, the ordering of ions leads to a small positive change in $\int (1-f(x))$ on increasing $c^{\mathrm{b}}_{\mathrm{NaCl}}$. 
This is sufficient to change the sign of $d\psi_i^{\mathrm{PZC}}/dc^{\mathrm{b}}_{\mathrm{NaCl}}$. 
It follows that, for small increases to $c^{\mathrm{b}}_{\mathrm{NaCl}}$, the resulting response to the potential of zero charge (PZC) can inform about the structure of the double layer.

\subsection{Ion Activities in the Double Layer}

\begin{figure*}[t]
\centering
  \includegraphics[width=0.7\linewidth]{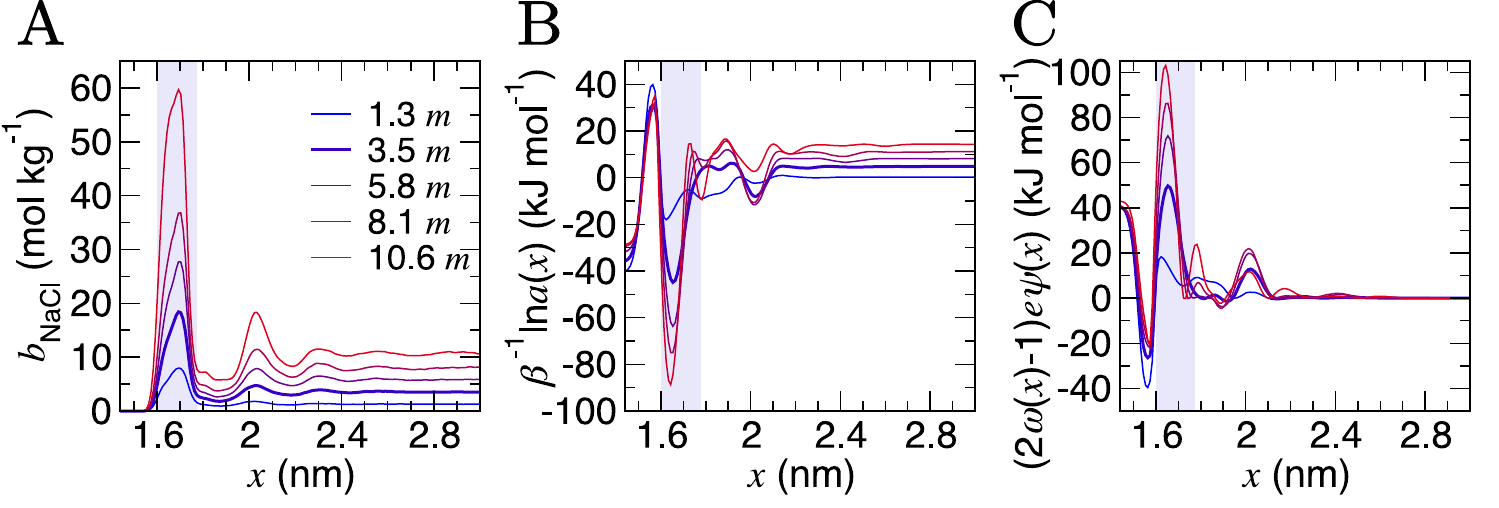}
   \caption[Activity]{Contributions to the electrochemical potential of electrolytes in the double layer region. A: NaCl molalities ($b_{\mathrm{NaCl}}^{\mathrm{b}}$) when the bulk mean molalities were as shown in the legend in units of $m$ (mol~kg$^{-1}$). B: The contribution to the mean ion molal electrochemical potential ($\tilde{\mu}$) from the first term in Equation \ref{eq:solution_equilibrium}. C: The contribution of the second term in Equation \ref{eq:solution_equilibrium} to $\tilde{\mu}$. The curve for the case when $b_{\mathrm{NaCl}}^{\mathrm{b}}=3.5$~mol~kg$^{-1}$ is highlighted by the bold line, and the shaded region indicates the position of the maximum peak in $b_{\mathrm{NaCl}}^{\mathrm{b}}$.}
  \label{fgr:activity}
\end{figure*}

The excess density of ions in the double layer has implications for the chemistry of this region. 
It is helpful to evaluate, therefore, contributions to the electrochemical potential, $\tilde{\mu_i}$, of ions, $i$, as a function of $x$:
\begin{equation}
\tilde{\mu_i}(x)=\mu^0+\beta^{-1}\ln{a_i(x)}+z_ie\psi(x)
\label{eq:electrochemicalpotdef}
\end{equation}
Here, $\mu^0$ is a reference chemical potential; $a_i(x)$ and $z_i$ are the activity and valency of species $i$, respectively; and, $\beta=1/k_{\mathrm{B}}T$, where $k_{\mathrm{B}}$ and $T$ are the Boltzmann constant and temperature, respectively. 
The total $\tilde{\mu}$ for NaCl(aq) in $x$ can be written as,
\begin{equation}
\tilde{\mu}(x)=\mu^0+\beta^{-1}\ln{a_\pm(x)}+(2\omega(x)-1)e\psi(x)
\label{eq:electrochemical_potential}
\end{equation}
where $\omega$ is the fraction of cations and $a_\pm(x)$ is a position-dependent average activity.  
We define $a_\pm(x)=a_{\mathrm{Na}}(x)^{\omega(x)}a_{\mathrm{Cl}}(x)^{1-\omega(x)}$, which converges to the mean ion activity, $a_\pm^{\mathrm{b}}=(a^{\mathrm{b}}_{\mathrm{Na}}a^{\mathrm{b}}_{\mathrm{Cl}})^{1/2}$, in the electroneutral bulk solution where $\omega(x)=0.5$.\cite{prausnitz_molecular_1999}
The electrochemical potential of solvated ions in the bulk, $\tilde{\mu}^{\mathrm{b}}$, is independent of $x$:
\begin{equation}
\tilde{\mu}^{\mathrm{b}}=\mu^0+\beta^{-1}\ln{a_\pm^{\mathrm{b}}}
\label{eq:electrochempot}
\end{equation}
At equilibrium, $a_\pm(x)$ and $\psi(x)$ are stationary, and the electrochemical potentials across the double layer and in the extended solution are equal:
\begin{equation}
\beta^{-1}\ln{a_\pm(x)}+(2\omega(x)-1)e\psi(x)=\beta^{-1}\ln{a_\pm^{\mathrm{b}}} 
\label{eq:solution_equilibrium}
\end{equation}

Based on fits to the chemical potentials of ions explicitly calculated in their own simulations and from other \textit{in silico} studies, \cite{moucka_molecular_2013,mester_mean_2015} Zimmerman et al. \cite{zimmermann_nucleation_2015,zimmermann_nacl_2018} provided an analytical model to calculate the chemical potential of solvated ions using the adopted force field:
\begin{equation}
\mu_{\mathrm{NaCl}} = \mu^0_{\mathrm{NaCl}} + 2 \beta^{-1}\mathrm{ln}\,b_{\mathrm{NaCl}} + 2 \beta^{-1}\mathrm{ln}\, \gamma_\pm
\label{eq:chempot-model}
\end{equation}
where $b$ is the mean molality in units of mol kg$^{-1}$, and $\gamma_\pm$ is the mean activity coefficient for ions; therefore, $a_\pm=b_{\mathrm{NaCl}}\gamma_\pm$, and,
\begin{equation}
\mathrm{log_{10}}\gamma_\pm = \frac{-A \sqrt{b_{\mathrm{NaCl}}}}{1+B \sqrt{b_{\mathrm{NaCl}}}}+C\,b_{\mathrm{NaCl}}
\label{eq:activity-coeff}
\end{equation}
where $A=0.568m^{-1/2}$, $B=1.17769m^{-1/2}$ and $C=0.177157m^{-1}$ (where $m=$~mol~kg$^{-1}$). 
Importantly, this model allows us to calculate ion activities at molalities far beyond the equilibrium saturation level of $3.7$~mol~kg$^{-1}$ \cite{moucka_molecular_2013,moucka_chemical_2015,mester_mean_2015,mester_temperature-dependent_2015,benavides_consensus_2016,espinosa_calculation_2016}. 
Values of $\mathrm{ln}\gamma_\pm$ and $\tilde{\mu}_{\mathrm{NaCl}}$ calculated using the above model are provided for the range of $b_{\mathrm{NaCl}}$ in Figure \ref{fgr:chempotmodel}. 

From C$\mu$MD simulations, the NaCl molality as a function of $x$ is calculated from atom density profiles
according to $(
n_{\mathrm{Na}}(x)n_{\mathrm{Cl}}(x))^{0.5}/(0.018n_{\mathrm{wat}}(x))$. 
Molalities in the bulk ($b^{\mathrm{b}}_{\mathrm{NaCl}}(x)$) were calculated from averages in the molality profiles in stable regions far from the interface. 
$b^{\mathrm{b}}_{\mathrm{NaCl}}(x)$ were substituted into Equations \ref{eq:activity-coeff} and \ref{eq:chempot-model} to calculate $(\mu_{\mathrm{NaCl}} - \mu^0_{\mathrm{NaCl}})/2$ which equals the right hand side of Equation \ref{eq:solution_equilibrium}. 
Note that this approach assumes equal contribution of cations and anions to the mean ion electrochemical potential. 
The C$\mu$MD simulation technique used here ensures accurate estimates of the bulk chemical potential of ions---within the adopted model in Equation \ref{eq:chempot-model}---where cation and anion concentrations must be uniformly equal within a small uncertainty.
%Bulk ion molalities relate to concentrations in the bulk according to $b_{\mathrm{NaCl}}^{\mathrm{b}}=0.03(c_{\mathrm{NaCl}}^{\mathrm{b}})^2+0.975(c_{\mathrm{NaCl}}^{\mathrm{b}})+0.036$. 

Analyses were performed using the density profiles in Figure \ref{fgr:density-profiles}, with Figure \ref{fgr:activity}~A providing the NaCl molality profiles for five systems. 
These reach a maximum at $x=1.6-1.7$~nm: the position close to the minimum following the first peak in cation concentration profiles. 
An increasing peak around $x=2$~nm matches with the increase in cation concentrations in this region at high concentrations. 
Counter-intuitively, the region in $x$ around the maximum  $b^{\mathrm{b}}_{\mathrm{NaCl}}(x)$ corresponds to a minimum in $\beta^{-1}\ln{a_\pm(x)}$ (see the shaded region Figure \ref{fgr:activity}~B).
This is due to a maximum in $(2\omega(x)-1)e\psi(x)$ as shown in Figure \ref{fgr:activity}~C. 
The fact that at $x=1.63$~nm, $\omega$ and $\psi$ are at a minimum, results in the large positive contribution to $\tilde{\mu}$ from this term in Equation \ref{eq:solution_equilibrium}. 
Beyond around 1~nm from the graphite surface, we find that the contribution of $(2\omega(x)-1)e\psi(x)$ to $\tilde{\mu}$ is zero, and in this region $\tilde{\mu}$ reduces to $\mu$. 

An important implication of the above observation is that the local ion molality (or concentration) at interfaces is not a good proxy for the electrochemical potential of ions. 
In systems where interfaces and extended liquid phases are in equilibrium, this is usually not problematic, due to the equality $\tilde{\mu}=\mu^{\mathrm{b}}$ across the boundary layer and into the bulk solution. Although any partial charge transfer of surface-bound ions should be considered. \cite{zhan_specific_2019,wang_demystifying_2021}
In some surface-driven processes at equilibrium, knowledge of the interfacial structure might still be essential to predict outcomes. 
For example, in processes like salt precipitation, the rates for nucleation are affected by the kinetic factors associated with the supply of ions to growing crystalline embryos. \cite{kashchiev_nucleation_2000}
Increased ion molalities close to the surface (being around five times the levels of the bulk, when the bulk molality equals the equilibrium saturation level of $3.7$~mol~kg$^{-1}$) likely mitigates the barriers to these processes.
In non-equilibrium processes, knowledge of both the local molality of solute species and the electric potential is essential to determine $\tilde{\mu}$.

\subsection{Ion Correlations and Diffusion}
\begin{figure*}[tb]
\centering
  \includegraphics[width=1.0\linewidth]{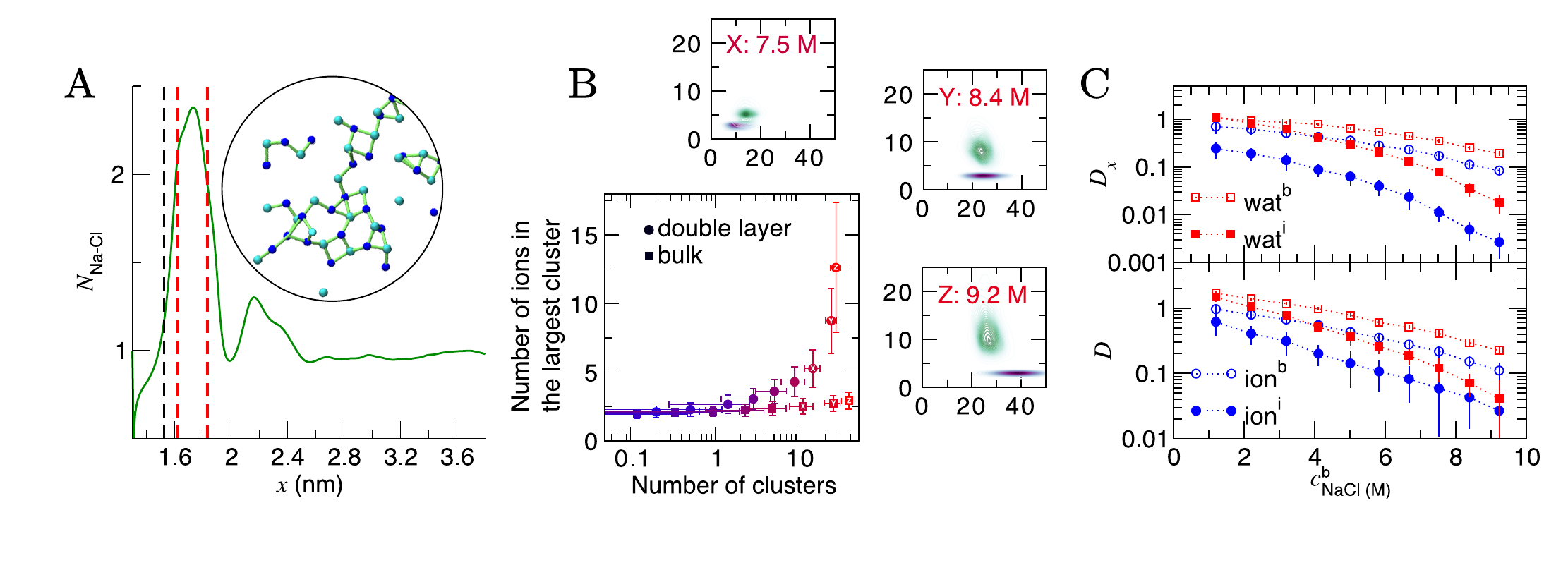}
   \caption[Diffusion]{Ion assembly and diffusion in the double layer. A: Average Na--Cl first-sphere coordination number ($N_{\mathrm{Na-Cl}}$) as a function of $x$ calculated from a C$\mu$MD simulation where $c^{\mathrm{b}}_{\mathrm{NaCl}}=9.2$~M. The black and red dashed lines indicate the maximum first Na and first two Cl densities in the concentration profiles for Na and Cl highlighted in Figure \ref{fgr:density-profiles}. Inset is a configuration of a large ionic network identified within the red dashed lines in $x$. The image is projected onto the $yz$ plane; blue and cyan spheres represent Na$^+$ and Cl$^-$ ions and the green lines highlight ions that are directly coordinated. B: Ion clusters observed in the bulk and double layer regions of simulations. The blue$\rightarrow{\mathrm{red}}$ colour scale in the main plot indicates increasing $c^{\mathrm{b}}_{\mathrm{NaCl}}$ from 1.2 to 9.2~M. The surrounding panels show purple and green probability densities in the same 2D space for clusters in the bulk and in the double layer, respectively, at the highest bulk concentrations shown by the legend. C: Diffusion coefficients, $D$, for ions and water in 0.4~nm regions at the interface and in the bulk region of C$\mu$MD simulations (indicated by the subscripts i and b in the legends). The top panel provides the $x$ component of $D$. Data have been scaled by $1\times 10^{-5}~\mathrm{cm}^2~\mathrm{s}^{-1}$ and error bars show uncertainties of one standard deviation.}
  \label{fgr:diffusion}
\end{figure*}

Simulations at the atomic level are perfectly suited to provide details regarding the collective arrangement and motion of ions in the liquid near the graphite surface. 
In this section, we characterise the species which emerge in the double layer at uncharged graphite surfaces and their diffusion.

\paragraph{Ion speciation.} We investigate ion speciation by calculating the average first shell coordination number, $N_{i-j}$, between atoms $i$ and $j$ described in SI Section \ref{sec:methods}. 
The average coordination numbers, $N_{\mathrm{Na-Cl}}=0.02 \pm 0.02$; $N_{\mathrm{Na-Ow}}=5.89 \pm 0.05$ and $N_{\mathrm{Cl-Ow}}=7.23 \pm 0.07$ were calculated for a 1~M NaCl(aq) bulk solution simulated for 10~ns. The structure of the solvated ions agrees well with other simulation and experimental studies. \cite{marcus_ionic_1988}
A majority of ions form solvent shared and solvent separated ion-pairs in the bulk (represented by the peaks at $r\approx 0.5$ and 0.7~nm in Figure \ref{fgr:rdfs-s}). 

Figure \ref{fgr:CNprofiles} provides the average $N$ as a function of $x$ evaluated using Gaussian kernel (with 0.03~nm bandwidth) probability densities. 
These indicate that the average $N_{\mathrm{Na-Ow}}(x)$ decreases by a value of one from the lowest to highest sampled concentrations in the bulk regions as $N_{\mathrm{Na-Cl}}(x)$ changes from zero to one, with these changes becoming significant when $c^{\mathrm{b}}_{\mathrm{NaCl}} \gtrsim 1$~M. 
Interestingly, $N_{\mathrm{Cl-Ow}}(x)$ is far less sensitive to changes in $c^{\mathrm{b}}_{\mathrm{NaCl}}$, which remain around the value identified in the bulk at 1~M over then entire bulk concentration range within the model.
The $N_{\mathrm{Na-Cl}}(x)$ coordination profiles in Figure \ref{fgr:CNprofiles}~A indicate a greater number of directly coordinated ions beyond the position of maximum densities in $x$ for the first ion layers. 
Essentially, the increased coordination occurs in regions of the double layer where there is a high density of both cations and anions. 
The maximum in $N_{\mathrm{Na-Cl}}(x)$ shifts to smaller values of $x$ and an additional peak emerges at $x \sim 2.2$~nm as crowding in the double layer increases. 
The features of these profiles roughly correspond to the profiles of $b^{\mathrm{b}}_{\mathrm{NaCl}}(x)$ in Figure \ref{fgr:activity}~A, which is a good indication that the association of ions in the double layer is due to increased ion densities. 

At the highest levels of $c^{\mathrm{b}}_{\mathrm{NaCl}}$, $N_{\mathrm{Na-Cl}} > 2$ at the maximum positioned at $x=1.75$~nm, as shown in Figure \ref{fgr:diffusion}~A. Concomitant changes occur to $N_{\mathrm{Na-Ow}}(x)$ in this region, with an additional minimum at $x=1.6$~nm.
The exceedingly high anion concentrations here affected the ability for water to fully solvate cations.
The increased Na--Cl coordination was due to the formation of many contact ion pairs that dynamically (dis)associate on the timescales of the simulations, leading to extended liquid-like networks of the type identified in the inset of Figure \ref{fgr:diffusion}~A. 
These structures begin to emerge at the graphite surface in significant numbers when the concentration of ions in the bulk exceeded approximately 5~M (i.e., beyond the nominal equilibrium saturation level for this force field). 
The networks are reminiscent of other liquid-like ionic networks identified in simulation studies, \cite{demichelis_stable_2011} which were suggested as precursors to crystalline phases. \cite{sebastiani_water_2017,smeets_classical_2017}

To analyse these structures further, we performed cluster analyses using the method of Tribello et al. \cite{tribello_analyzing_2017} (see SI Section \ref{sec:methods} for details). 
Figure \ref{fgr:diffusion}~B indicates that the effect of increasing $c^{\mathrm{b}}_{\mathrm{NaCl}}(x)$ is to increase the number of ion clusters (defined as species containing more than two ions in direct contact) in the bulk and within the double layer. 
In the bulk, these clusters contain a maximum of three ions. 
In the double layer, however, the clusters contained more than ten ions at the highest $c^{\mathrm{b}}_{\mathrm{NaCl}}(x)$, with a wide distribution in the size of the largest clusters due to the rapid time evolution of ion--ion correlations. 
Both the size and geometry of the networks rapidly changed over several nanoseconds of simulation and exchange of ions with the surrounding solution occurred. 
Similar liquid-like NaCl clusters were identified in simulations beyond the limit of solution stability (15 mol~kg$^{-1}$) in the bulk, where a change in the mechanism for salt precipitation occurs. \cite{jiang_nucleation_2019}
Some experimental studies posit the existence of NaCl clusters even at moderate saturation levels. \cite{georgalis_cluster_2000}
Further studies are now needed to explore the role that liquid-like clusters play on the nucleation of NaCl at interfaces.

\paragraph{Diffusion in solution.} The diffusion coefficients, $D$, for ions were measured using the Einstein relation, described in detail in SI Section \ref{sec:methods}. 
For reference, the average $D$ for ions  measured in simulations of bulk of 1~M NaCl(aq) was $1.14 \pm 0.05~\times 10^{-5}~\mathrm{cm}^2~\mathrm{s}^{-1}$.

$D$ and $D_x$ as a function of $x$ are provided in Figure \ref{fgr:diffusion-all}; these indicate that the surface decreases the diffusion of ions and water molecules within the double layer. 
The decrease in both $D(x)$ and $D_x(x)$ on approach to the substrate is monotonic; hence, the diffusion coefficients closest to the graphite surface and in the bulk region have been plotted as a function of $c^{\mathrm{b}}_{\mathrm{NaCl}}$ in Figure \ref{fgr:diffusion}~C for simulations sampling the higher end of the entire concentration range.
The values of $D$ were found to decay following an approximately exponential trend: $D=D_0e^{-\lambda c^{\mathrm{b}}}$, where $\lambda$ is the so-called decay constant. 
In the bulk, $D_0$ were 1.344 and $2.396 \times 10^{-5}~\mathrm{cm}^2~\mathrm{s}^{-1}$ and $\lambda$ were $-0.236$ and $-0.386$ for ions and water, respectively. 
A more negative $\lambda$ for water indicates that increasing ion concentrations retards the mobility of the solvent molecules moreso than solute ions. 
At the interface, however, $\lambda$ for ions was $-0.335$ ($D_0=0.819$): more negative than  for water ($\lambda=-0.221$; $D_0=2.248$), which is most likely due to the increased concentration of ions in this region and the changes to the speciation of ions, discussed above. 

Around 1~M, $D^{\mathrm{i}}/D^{\mathrm{b}}=0.64$ and 0.87 for ions and water, respectively (where $D^{\mathrm{i}}$ and $D^{\mathrm{b}}$ are the diffusion coefficients close to the graphite surface and in the bulk). 
At the highest concentrations sampled, $D^{\mathrm{i}}/D^{\mathrm{b}}=0.24$ and 0.18 (for ions and water, respectively). 
The arresting of particle mobilites is largely due to decreased diffusion perpendicular to the interface. The top panel in Figure \ref{fgr:diffusion} provides the $x$ component of $D$ for water and ions at the interface and in the bulk. 
At 1~M, $D_x^{\mathrm{i}}/D_x^{\mathrm{b}}=0.35$ for ions and this reduces to 0.03 at the maximum bulk concentration. 
This reflects the high charge densities close to the graphite surface. 
In contrast, $D_x^{\mathrm{i}}$ for water molecules is unchanged compared to $D_x^{\mathrm{b}}$ at the lowest concentration, but beyond 9~M, $D_x^{\mathrm{i}}/D_x^{\mathrm{b}}=0.09$, and the high salinity interface retards the mobility of water molecules in $x$ nearly as significantly as for ions.

Ion transport properties in the double layer are often assumed to match with those in the bulk, e.g., when calculating $\zeta$-potentials using electrokinetic flow apparatus.
Diffusion coefficients for ions in solution near the graphite surface on the order $1\times 10^{-7}-1\times 10^{-5}~\mathrm{cm}^2~\mathrm{s}^{-1}$ indicate an increased viscosity in the double layer caused by the changing solution densities in this region.
While direct coordination of cations to the graphite was evident, no specific surface-site binding was identified, and diffusion was particularly limited perpendicular to the graphite surface. 
This picture is arguably consistent with the idea of a `dynamic Stern layer'. \cite{rubio-hernandez_primary_2004,dopke_importance_2021}
However, this term is unhelpful,\cite{schmickler_double_2020} failing to recognise the dynamic equilibrium between ions in the first and adjacent solution layers.
No clear boundary (slipping plane) between the diffusion of ions in a specifically adsorbed layer at the surface and in the diffuse region can be identified from $D$ or $D_x$ in Figure \ref{fgr:diffusion-all}.
We refer the reader to a recent monograph by D\"opke and Hartkamp, \cite{dopke_importance_2021} where these effects are discussed in the context of  electrokinetic phenomena.

\subsection{Graphite with Applied Electric Charge}
\label{sec:charged}
\begin{figure*}[t]
\centering
  \includegraphics[width=.9\linewidth]{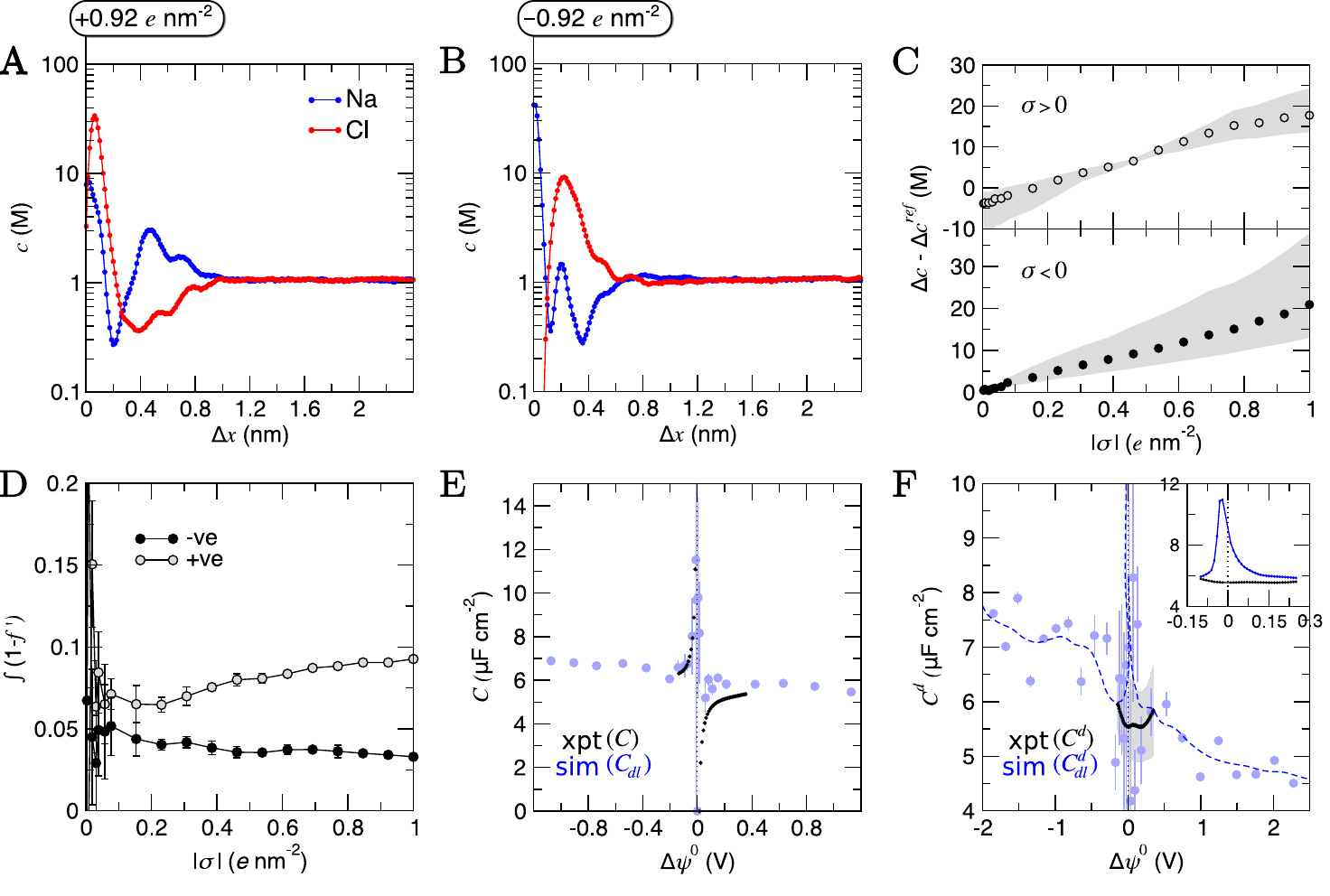}
   \caption[Surface Charge Effects]{The effects of applied surface charges on the double layer. A and B provide ion concentrations as a function of distance from the electrode (positioned at $x=0$) measured in C$\mu$MD simulations when $|\sigma|=\pm0.92e$~nm$^{-2}$, where $\sigma$ is the surface charge density. C provides the difference in the maximum ion concentrations ($\Delta c$) at the electrode as a function of $\sigma$. Equivalent concentration differences, from distributions when no surface charge is applied ($\Delta c^{ref}$), were first subtracted, and shaded areas show uncertainties in the MD data due to changes using a smoothing window of $0.2 \pm 0.1$~nm. D shows the integrals of $1-f'$, where $f'$ is the electrode screening factor, calculated for all atoms in the double layer region in solutions. E provides the capacitance ($C$) as a function of the potential difference across the double layer after subtracting the potential of zero charge ($\Delta \psi^0$). Experimental data are calculated from the integral of the differential capacitance ($C^d$) data in panel F for the 1~M case. F provides the measured $C^d$ from experiments and the evaluated double layer capacitance from simulations ($C^d_{dl}$) as a function of $\Delta \psi^0$. The dashed blue line is a moving average over three data points (in blue) at positive/negative $\Delta \psi^0$. This is shown more clearly in the inset of F over the range of potential differences used to measure $C^d$ (also inset and shown by the black data points). Where shown, error bars highlight the standard error of the mean from multiple trajectory window analyses (otherwise statistical uncertainties are of the size or smaller than the data points), with the grey region in F showing the uncertainty in the mean $C^d$ from three repeat experiments. In E and F, data have been truncated for ease of comparison between simulations and experiments.}
  \label{fgr:electrode-main}
\end{figure*}

To consider the effect of applied electric fields, charges were uniformly distributed to the outermost carbon atoms at the graphite basal plane, discussed in detail in SI Section \ref{sec:methods}. 
Negatively and positively charged surfaces were simulated with charge densities, $\sigma$, in the range $| \sigma |=0.004$--$1~e~\mathrm{nm}^2$ (see SI Table \ref{tab:conc}).
Our approach neglects the electronic response of the graphite to charging the material, which must be considered when comparing to experiments.
This is reasonable, considering that DFT calculations indicate that applying an electric field to graphite induces equal and opposite net excess $\sigma$, centered close to the edges of a graphite slab, perpendicular to the field direction. \cite{luque_electric_2012}

The concentration of ions beyond the double layer was maintained in C$\mu$MD simulations (see Figure \ref{fgr:electrode-CR-concn}) where $c^{\mathrm{b}}_{\mathrm{NaCl}}=1.05 \pm 0.03$~M.
Assuming Poisson-Boltzmann behaviour, Grahame's Equation (see Equation \ref{eqn:grahame}) provides a direct relationship between (the effective) $\sigma$ and the potential change across the double layer, $\Delta \psi$.
Even within the monotonic regime, Poisson-Boltzmann approximations fail to accurately predict $\sigma$. \cite{attard_ion_1995,bazant_towards_2009,dopke_importance_2021} 
These models are, therefore, unhelpful, particularly to determine the interfacial properties of NaCl(aq) on graphite at technologically-relevant electrolyte concentrations, as discussed below.

\paragraph{Structural asymmetries in the double layer.}
Ion number densities as a function of $x$ in Figure \ref{fgr:electrode-atomdensities} show that when increasingly positive surface charges are applied, Cl$^-$ ions were pulled closer to the graphite surface in a less diffuse anion layer. 
Concomitantly, the maximum density in the first cation layer decreased and was shifted further away from the surface.
Despite this, a more compact double layer emerges due to increased cation densities in a second cation layer, shown most clearly at high values of $\sigma$ in Figure \ref{fgr:electrode-main}~A.
When the surface charge was made incrementally negative, the first two cation layers both increased in density and peaks in $n_{\mathrm{Na}}(x)$ appear sharper. 
The maximum in the anion density decreased and was shifted away from the graphite, although the anion layer remained diffuse, with the density of the tails in the distribution actually increasing (see Figure \ref{fgr:electrode-main}~B and Figure \ref{fgr:electrode-atomdensities}). 

Structural changes were observed in the solvent layers at the interface; these were particularly significant when $\sigma$ was large, in line with simulations elsewhere studying electrified planar interfaces.  \cite{goldsmith_effects_2021}
Figure \ref{fgr:electrode-waterdens-comp} shows that when $\sigma$ was large and positive, water oxygen atoms were pulled closer to the surface and peaks in $n_{\mathrm{Hw}}(x)$ appear sharper.
More interestingly, when $\sigma$ was more negative than approximately $-0.5~e~\mathrm{nm}^2$, a restructuring of water molecules was apparent at the interface, with a splitting of the first peak in $n_{\mathrm{Hw}}(x)$, as solvent molecules arrange their hydrogen atoms towards and away from the graphite surface.

Changes to the excess ion number densities ($n_i-n_i^{ref}$, where $n_i^{ref}$ are the densities when $\sigma=0$) as a function of $\sigma$ are reported in Figure \ref{fgr:electrode-residual-densities}. 
These highlight that the double layer undergoes an asymmetric enrichment of charge-balancing counter-ions and depletion of co-ions. 
The asymmetries can be quantitatively evaluated by considering the excess counter-ion concentration at the charged graphite surface. 
This was done by taking the difference in ion concentrations, $\Delta c =c_A - c_B$ (where $A$ is the counter-ion type and $B$ is co-ion), from the maximum in concentration profiles where $\Delta x<1.5$~nm in the MD data.
Figure \ref{fgr:electrode-main}~C shows that $\Delta c$ increases monotonically from a value of zero when $\sigma=0$; this is qualitatively consistent with the predictions of the GCS model.
At the positively charged surface, $\Delta c$ is negative when $\sigma$ is small.
The repulsion of cations in the first ion layer (beyond the electrode) leads to a small decrease in the maximum anion concentrations in the first anion layer and a small increase in the maximum concentrations in the second cation layer beyond the surface (see Figure \ref{fgr:electrode-residual-densities}).
These observations are beyond the predictions of simple mean-field models and are due to the complex interplay of charge screening and volumetric constraints due to finite ion size effects at intermediate--high bulk concentrations.

\paragraph{Electrode charge screening.}
Asymmetries in the shape of the excess solution charge densities (defined as $\rho^0(x)=\rho(x) - \rho^{ref}(x)$, where $\rho^{ref}$ is the $\sigma=0$ case, reported in Figure \ref{fgr:electrode-residual-electrics}) appear in the first ion layer, with a doublet peak emerging at the negatively charged surface.
Despite this, the features of the resulting E$^0(x)$ and $\psi^0(x)$ curves (calculated using the Poisson Equation, and also provided in Figure \ref{fgr:electrode-residual-electrics}) are largely symmetrical when comparing the oppositely charged surfaces.
Differences occur in the amplitudes of the fluctuations in these curves. 
Figure \ref{fgr:cap-deriv}~A highlights a greater change to the electric potential across the double layer region, $\Delta \psi^0$, in response to applying positive charges (cf. negative charges) at the graphite surface.
The potential difference diverges when  $|\sigma|=0.1~e~\mathrm{nm}^2$.

To further understand the above divergence, we calculated an electrode screening factor:
\begin{equation}
f'(x')=\frac{\int_0^{x'} (\rho_{\mathrm{NaCl}}(x)-\rho_{\mathrm{NaCl}}^{ref}(x)) dx}{-\sigma}
\label{eq:electrodescreeningfactor}
\end{equation}
Figure \ref{fgr:electrode-screening-curves} shows that, considering only ions in solution, over-screening occurs at the positively charged surface due to the high density of anions in the first solution layer at the highest $\sigma$.
$f'$ then decays until it converges to a constant value far from the graphite surface. Small changes to $\sigma$ affect the maximum in $f'(x)$ and, therefore, the gradient in the screening profiles as $f'$ converge to their bulk values. 
On the negatively charged surface, however, there is an initial over-screening that is compensated by charges in an adjacent solution layer, leading to a minimum in $f'$ around 1.6~nm. 
We calculated $\int_0^{x'}(1-f'(x))$ (after ensuring that all $f'$ profiles converge to a value of one). 
The resulting curves at the highest values of applied potential are provided in Figure \ref{fgr:electrode-screening-curves}. 
A positive (negative) shift in the converged value of $\int_0^{x'}(1-f'(x))$ is found at the positively (negatively) charged graphite surface. 
This is consistent with a more diffuse screening of the positively charged surface.
When all solution atoms were included in the analyses, the same trends were observed, as highlighted by Figure \ref{fgr:electrode-main}~D; although the data here are noisy due to fluctuations in the water density profiles, particularly for small values of $|\sigma|$. 
These changes correlate with the changes to $|\Delta \psi^0|$, presented in panel D of Figure \ref{fgr:cap-deriv}~A. 

The asymmetric accumulation of ions manifests in a greater capacity, $C=\sigma/\Delta \psi^0$, for the surface to store ionic charge when negative potentials are applied. 
Figure \ref{fgr:electrode-main}~E shows that maximum and minimum $C$ occur at small negative and positive values of $\Delta \psi^0$.
This is in very good agreement with the experimental $C$ values (also show in Figure \ref{fgr:electrode-main}~E) calculated from the integral of $C^d(\Delta \psi^0)$ measured for the HOPG--NaCl(aq) system when $c^{\mathrm{b}}_{\mathrm{NaCl}}=1$~M (described in SI \ref{sec:methods}). 
The asymptotic behaviour of $C(\Delta \psi^0)$ are opposite to the trends at positive/negative potential observed in simulation studies of molten LiCl at atomically flat electrodes; however, these electrodes are not related to any specific material. \cite{vatamanu_molecular_2010}
When $\Delta \psi^0$ was highly negative, $C$ is up to 2~\textmu{F}~cm$^{-2}$ greater than when $\Delta \psi^0$ was large and positive. 

Despite the close agreement between the simulated and experimentally determined values of $C$, the differential capacitance data ($C^d$; provided in Figure \ref{fgr:electrode-main}~F) show deviations around $\Delta \psi^0=0$.
Here, the mean simulated differential capacitance reaches a maximum around 11~\textmu{F}~cm$^{-2}$ (shown inset of Figure \ref{fgr:electrode-main}~F, with the average over the full range of potential difference sampled in the experiments being 6.1$\pm$1.2 \textmu{F}~cm$^{-2}$ from simulations). 
Our simulations, however, only evaluate the contribution to the series of differential capacitance, $1/C^d=1/C^d_{dl}+1/C^d_q$, associated with the double layer response to the applied charge ($C^d_{dl}$); they neglect any contribution to $C^d$ due to the (re)distribution of the graphite electron density of states associated to charging ($C^d_q$), which was postulated in early studies to dominate $C^d$. \cite{gerischer_density_1987,randin_differential_1971}
Our results, therefore, indicate that both $C^d_{dl}$ and $C^d_q$ contribute to the measured $C^d$ at the chosen electrolyte concentration around $\Delta \psi^0=0$.
This was confirmed by evaluating $C^d_q$ explicitly using $1/C^d_q =  1/C^d−1/C^d_{dl}$, as shown in Figure \ref{fgr:cap-comp}, which indicates that $C^d_q$ is of the same magnitude as $C^d_{dl}$, on average, around $\Delta \psi^0=0$.

Increasing the potential difference results in rapidly increasing $C^d_q$, with a larger magnitude in the gradient ($dC^d_q/d\Delta \psi^0$) on the anodic branch of the $C^d_q(\Delta \psi^0)$ curve.
This means that, at large values of $\sigma$, $C^d_{dl}$ dominates the measured $C^d$ at graphite.
Indeed, the shape of $C^d_q(\Delta \psi^0)$ matches well to experimental measurements of the `quantum capacitance' at graphene electrodes, although the size of $C^d_q$ is an order of magnitude greater here at similarly large values of applied potential. \cite{xia_measurement_2009}
This is due to the proportional increase in the electrode (electron) density of states per unit surface area that occurs as the number of graphene layers are increased, as determined in recent simulations at the DFT level. \cite{zhan_quantum_2015}
The DFT calculations show that a $C^d_q$ of around 10 \textmu{F}~cm$^{-2}$ (at the $C^d_q$ minimum) in 6--layer graphene is increased by an order of magnitude when $|\Delta \psi^0| \approx 0.5$~V, in close agreement to the evaluated $C^d_q$ here.

Early models used to explain the $C^d$ at graphite implicitly assumed monotonic behaviour in the concentration profiles of charges in solution adjacent to the electrode and an absence of specific adsorption, through their use of the GCS model when apportioning contributions to $C^d$. \cite{gerischer_density_1987,randin_differential_1971,luque_electric_2012} 
This was motivated by that fact that the minimum $C^d$ of graphite is approximately an order of magnitude smaller than the $C^d$ measured for metal electrodes at comparable conditions.
It was, therefore, implicitly assumed that $C^d$ was dominated by $C^d_q$. \cite{gerischer_density_1987,randin_differential_1971,luque_electric_2012} 
Outside of the GCS model, however, there is no reason to assume that graphite electrodes accumulate solution charge in the same way as metal electrodes, such that they should have similar magnitudes of $C^d_{dl}$; indeed, this was briefly considered early on.\cite{randin_differential_1972,gerischer_density_1987}

Our estimates of $C^d_{dl}$ are, to some degree, force field dependent.
At the concentrations adopted, however, it is likely that other force fields, which capture (at least implicitly) the polarisation at the graphite surface, will determine a similar value for $C^d_{dl}$ due to the partial `saturation' of the multi-layered double layer structure at high concentrations.
In addition, the possible adsorption of airborne contaminants \cite{zou_investigation_2016} at the surface, and the effect of steps and other surface defects in HOPG should be considered when attempting any comparison between simulations and experiments.
Nonetheless, our results are consistent with a dependence of $C^d$ on the solution composition, reported here and elsewhere\cite{iamprasertkun_capacitance_2019,iamprasertkun_understanding_2020}.
Asymmetries in the tails of the $C^d$ curves are accentuated in the simulation results, and these are apparent also in experiments---highlighted more clearly by the changing gradients of the $C^d(\Delta \psi^0)$ curve in Figure \ref{fgr:cap-deriv}~B. 
This is a common feature in studies of ionic liquids, \cite{fedorov_ionic_2014} where the anisotropic, bulky charge carriers saturate opposing surface charge densities differently.

\section{Conclusions}
C$\mu$MD simulations were applied to simulate the graphite--NaCl(aq) interface in equilibrium with constant density, electroneutral bulk solutions sampling a range of electrolyte concentrations (0.2--9.2~M).
Na$^+$ accumulation at the graphite surface in a dynamically adsorbed first layer (as confirmed by analyses of ion diffusion coefficients) results in an effective positive surface charge density that is charge-compensated by the accumulation of Cl$^-$ ions. 
At the lowest bulk electrolyte concentrations, the concentration of anions decreases exponentially (exponential fit $R^2=0.94$ at 0.2~M), with the double layer thickness shrinking with increasing concentration. 
Above 0.6~M, however, a transition in the screening behaviour occurs, with  alternating layers of cations and anions (up to four--five solvent layers in extent) forming before the effective surface charge is neutralised, leading to an increasing double layer size with concentration.
The crowding of ions, and increasing over-screening in the double layer region, manifests in a reduction to the change in the potential of zero charge with incremental changes to the bulk electrolyte concentration, which was confirmed experimentally.

Charging the graphite surface, when $c^{\mathrm{b}}_{\mathrm{NaCl}}\approx1$~M, allowed evaluation of the double layer differential capacitance ($C^d$).
That the average over the minimum region in simulations (6.1$\pm$1.2 \textmu{F}~cm$^{-2}$) and experiments (5.54$\pm$0.60 \textmu{F}~cm$^{-2}$) were of the same order of magnitude highlights a problem with previous analyses of the double layer capacitance ($C^d_{dl}$) of the graphite--electrolyte interface, which assumed that a GCS model was applicable. \cite{randin_differential_1971,gerischer_density_1987}
We emphasise that parallels cannot be drawn between the classical picture of the compact layer capacitance at metal electrodes and the graphite electrode; this led us to reconsider the relative contributions to the total $C^d$.
Estimates of $C^d$ associated with the response of the graphite electronic structure to charging ($C^d_q$), using our simulated values for $C^d_{dl}$ and total measured $C^d$, shows that $C^d_q$ is of a similar magnitude to $C^d_{dl}$ around $\Delta \psi^0 = 0$ (around 10 \textmu{F}~cm$^{-2}$), increasing substantially with the value of applied potential.
This is in line with recent simulations of many-layer graphene.\cite{zhan_quantum_2015} 
The significance of this result is that $C^d$ (at moderate-to-high electrolyte concentrations) does not simply report on the density of states of the graphite material, as has become the accepted norm in much of the electrochemistry literature.

Evaluation of interfacial molalities showed increases of up to six times the levels of the bulk solution in regions confined to the double layer, where the solution mass densities were up to three times the levels of the bulk (1030~kg~m$^{-3}$ at 1~M).
Despite this, the mean ion activities are reduced due to the large local electric fields present, emphasising the non-ideality of this region. 
Cluster analyses reveal the presence of enhanced clustering at concentrations above 5~M, which has implications for the rates of surface-driven reactions.
The clusters are reminiscent of those identified in simulations at the limit of solution stability in bulk solutions ($\sim 15$~mol~kg$^{-1}$), where phase separation becomes spontaneous. \cite{jiang_nucleation_2019} 
Given that the clusters emerge in the metastable solution (i.e., beyond the nominal solubility of NaCl in the bulk and below the limit of solution stability), further investigation is required to determine their role in salt precipitation facilitated by surfaces.

\section{Acknowledgements}
The authors acknowledge funding from an EPSRC Programme Grant (Grant EP/R018820/1). The authors acknowledge the use of the UCL Myriad High Throughput Computing Facility (Myriad@UCL), and associated support services, in the completion of this work.

%\bibliography{REECl_paper_SS} 
\bibliographystyle{achemso}
\bibliography{graphite-nacl}
\clearpage

% SUPPORTING INFO
\pagenumbering{arabic}
\appendix
\setcounter{figure}{0} 
\setcounter{table}{0} 
\setcounter{equation}{0} 
\renewcommand\thefigure{S\arabic{figure}}
\renewcommand\thetable{S\arabic{table}}

\newpage
\onecolumn
{
\centering
\section*{Electrochemistry, Ion Adsorption and Dynamics in the Double Layer: A Study of NaCl(aq) on Graphite}

\vspace{0.5 cm}
{\Large\emph{\emph{Supporting Information}}} 
\vspace{1 cm}

{\large Aaron R. Finney$^{\dagger,\ast}$, Ian J. McPherson,$^{\ddagger}$ Patrick R. Unwin,$^{\ddagger}$ and Matteo Salvalaglio$^{\dagger,\ast}$}
\newline

$^\dagger$\textit{Thomas Young Centre and Department of Chemical Engineering, University College London, London WC1E~7JE, United Kingdom}

$^\ddagger$\textit{Department of Chemistry, University of Warwick, Coventry, CV4~7AL, United Kingdom}
\newline

$^\ast$To whom correspondence may be addressed: a.finney@ucl.ac.uk; m.salvalaglio@ucl.ac.uk

}
\vspace{1 cm}
\etocdepthtag.toc{mtappendix}
\etocsettagdepth{mtchapter}{none}
\etocsettagdepth{mtappendix}{paragraph}
\etocsetnexttocdepth {paragraph}

\tableofcontents

\clearpage
\section{Materials and Methods}
\label{sec:methods}
\subsection{Simulation Details}
\paragraph{Simulation setup.}
A $5.4 \times 5.5 \times 2.7$~nm graphite supercell was created from the unit cell structure elucidated by Trucano and Chen,\citeSI{trucano_structure_1975} after first transforming the unit cell to an orthorhombic geometry. 
The resulting structure contained eight layers of graphene positioned perpendicular to the simulation cell $x$ axis. 
Two different systems were prepared, namely system A and B. 
In system A, 11,389 water molecules, 418 Na$^+$ and 418 Cl$^-$ ions were added to the simulation cell above and below the graphite with respect to $x$. 
With carbon atoms restrained to their crystalline lattice positions, a molecular dynamics (MD) simulation was performed for 0.2~ns at 298~K and 1~bar to relax the solution and equilibrate the simulation cell volume. 
The resulting solution mass density was 1.05~g~cm$^{-3}$ with an ion molality, $b($NaCl$)=2$~mol~kg$^{-1}$. 
System B was prepared following the same protocol using 1,672 NaCl and 13,819 water molecules, leading to a solution mass density of $1.19$~g~cm$^{-3}$ and ion molality $b($NaCl$)=6.7$~mol~kg$^{-1}$.

MD simulations were performed using the leapfrog time integration algorithm with a 2~fs timestep in the GROMACS 2018.6 package \citeSI{hess1_gromacs_2008}. 
The temperature and pressure of the system were held constant within statistical fluctuations using the Bussi-Donadio-Parrinello thermostat \citeSI{bussi_canonical_2007} and the barostat of Berendsen et al. \citeSI{berendsen_molecular_1984} 
The internal degrees of freedom of water molecules were constrained using the LINCS algorithm \citeSI{hess_lincs:_1997}. 
Smooth particle mesh Ewald summation \citeSI{essmann_smooth_1995} was adopted to compute atomic energies and forces arising from electrostatic interactions, where the real-space contributions were computed for atoms within 0.9~nm. 
Lennard-Jones interactions were truncated at 0.9~nm with a dispersion correction added to the energies of short-range intermolecular interactions. 
Periodic boundaries were imposed in three-dimensional Euclidean space. 

To create the initial configurations for C$\mu$MD simulations \citeSI{perego1_molecular_2015}---where an internal, high salinity region far from the graphite surface is used as a reservoir to control the concentration in the system---the distances between all ions and graphite carbon atoms were restrained to 6~nm using a harmonic bias potential with a force constant of $3 \times 10^5$~kJ~mol$^{-1}$. 
Systems A and B were simulated for 0.5~ns in the canonical ensemble under the effect of this restraint, imposed using the PLUMED plugin (version 2.5) \citeSI{tribello1_plumed_2014}.

\paragraph{C$\mu$MD.}

\begin{figure}[h]
\centering
  \includegraphics[width=0.5\linewidth]{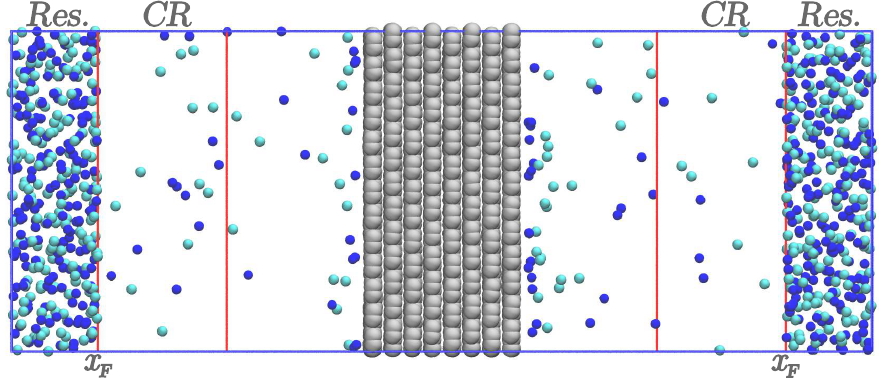}
  \caption[C$\mu$MD simulations in this work]{An example configuration (projected along $x$ and $z$ Cartesian coordinates) taken from a C$\mu$MD simulation with target 0.06~nm$^{-3}$ Na$^+$ and Cl$^{-}$ number densities in the control region, CR. Sodium and chloride ions are shown as blue and cyan spheres, respectively. Carbon atoms of graphite are shown as grey spheres. Water molecules are omitted for clarity. Red lines mark the boundaries for the CR and reservoir (Res.) regions.}   
  \label{fgr:cmu-snap}
\end{figure}

C$\mu$MD \citeSI{perego1_molecular_2015} was adopted to maintain a constant number density of ions, $n^t$, in the control region (CR) solutions, relatively far from the graphite/solution interface (see Figure \ref{fgr:cmu-snap}), in simulations where the total volume was held constant. 
The number density is controlled by applying a continuous, external force, $F_i^{\mu}$, to ions at $x_F$: the boundary between the CR and the reservoir. 
When the simulation is initiated, ions from the reservoir diffuse towards the interface and $F^{\mu}$ acts as a semi-permeable membrane to maintain a constant, predefined value of cation and anion number densities in the CR:
\begin{equation}
    F_i^{\mu}(x)=k_i(n_i^{\mathrm{CR}}-n_i^{t})G(x)
\end{equation}
Here, $n_i^{\mathrm{CR}}$ is the ion $i$ number density in the CR. 
A continuous force is applied to ions at a fixed distance, $x_F$, from the midpoint ($x_0$) of the line spanning the $x$ axis:
\begin{equation}
    G(x)=\frac{1}{4 \omega} \left[ 1 + \mathrm{cosh} \left( \frac{x-x_F}{\omega} \right) \right]^{-1}
    \label{eqn:bellfunction}
\end{equation}
In these simulations, the graphite--solution interface does not change its position in time i.e., the position $x_F$ is time-independent.  
Note that an alternative approach, where the position at which the force is applied evolves during the simulation, has been adopted in C$\mu$MD studies in which a phase transition takes place at the solid/liquid interface which evolves in time\citeSI{perego1_molecular_2015}. 
In Equation \ref{eqn:bellfunction}, $\omega$ controls the width of the force region acting on particles close to $x_F$. 
In the C$\mu$MD simulations here, $\omega$ was 0.01\% of the total size of $x$ and $k=2 \times 10^5$~kJ~mol$^{-1}$; hence, the force is localised to a very narrow region in $x$. 
The size of the CR in $x$ was 2.2~nm and $x_F=x_0\pm 3.7$~nm. 

C$\mu$MD simulations were performed for 100~ns setting $n^t$ to 0.06022--0.6022 nm$^{-3}$ in 0.06022 nm$^{-3}$ increments in system A and $n^t=$ 0.6022--6.022 in 0.6022 nm$^{-3}$ in system B.
The densities correspond to target ion molar concentrations, $c^t=n^t \times 10^{24}/N_{\mathrm{A}}$ (where $N_{\mathrm{A}}$ is Avogadro's constant), of 0.1--10 M (mol dm$^{-3}$). 
Typically, 20--35~ns of simulation time was required to reach a steady state in ion concentration profiles; we therefore utilised the final 50~ns of simulation trajectories in any analyses.

\paragraph{Force field.}

NaCl(aq) was simulated using the Joung and Cheatham force field \citeSI{joung_determination_2008} which adopts the SPC/E model \citeSI{berendsen_missing_1987} for water. 
This model reproduces reasonably well the thermodynamics of NaCl in water.\citeSI{aragones_solubility_2012} 
Graphite C--C atomic interactions were modelled using the OPLS/AA force field \citeSI{jorgensen_development_1996}. 
C--water intermolecular interactions were modelled using the atom pair potential provided by Wu and Aluru \citeSI{wu_graphitic_2013} based on fitting to water adsorption energies from random phase approximation calculations\citeSI{ma_adsorption_2011}. 
This model predicts a water contact angle of $42^{\circ}$, which the authors highlight compares favourably to experimental measurements\citeSI{schrader_ultrahigh_1975,prydatko_contact_2018}. 
Other experiments\citeSI{li_effect_2013} suggest a less hydrophilic wetting angle of $64^{\circ}$ on freshly cleaved HOPG that increases over time due to contamination in the atmosphere. 
The model reportedly predicts well the radial breathing mode frequency of single walled carbon nanotubes. \citeSI{wu_graphitic_2013}
Finally, (graphite) C--Na$^+$ and C--Cl$^-$ intermolecular interactions were modelled using the atom pair potentials from Williams et al.\citeSI{williams1_effective_2017}. 
These potentials were fitted to the energies of interaction of ions with aromatic, planar molecules containing 54 carbon atoms and with solvent implicitly modelled using a conductor-like polarisable continuum. 
Importantly, there is a self-consistency associated with different components of the force field; for example, the ion--water model adopted in the derivation of C--ion potentials was the Joung and Cheatham force field and SPC/E water was used in the fitting of C--water interactions.

\paragraph{Applied surface charge.}

Simulations to explore graphite surface charge effects were performed where, unless otherwise stated, the simulation input parameters were the same as for C$\mu$MD simulations in the absence of a graphite surface charge.
In total, 24, 100~ns simulations were initiated from a C$\mu$MD simulation in the absence of a graphite surface charge and with $c^{\mathrm{b}}_{\mathrm{NaCl}}=1.05 \pm 0.03$~M. 
Surface charges were applied uniformly to the outermost carbon graphite plane. 
The sign of the charge was opposite on different sides of the basal planes of the graphite slab; hence surfaces with a negative and positive surface charge density, $\sigma$, were exposed to NaCl(aq).
Any induced dipole moment in $x$ from the additional surface charges can be screened by the redistribution of charges within the internal reservoir. 
This approach to exploring electrified interfaces in simulations is somewhat simplistic; nonetheless, it allows for a direct appraisal of the GCS model.
More sophisticated simulation methods are available to e.g., control the electric potential \citeSI{vatamanu1_molecular_2010} or to capture any dynamic electronic effects at the interface \citeSI{elliott1_qmmd_2020,goldsmith1_effects_2021}; however, these are costly and do not ensure a constant concentration of cations and anions in the bulk.

\begin{figure}[ht]
\centering
  \includegraphics[width=0.45\linewidth]{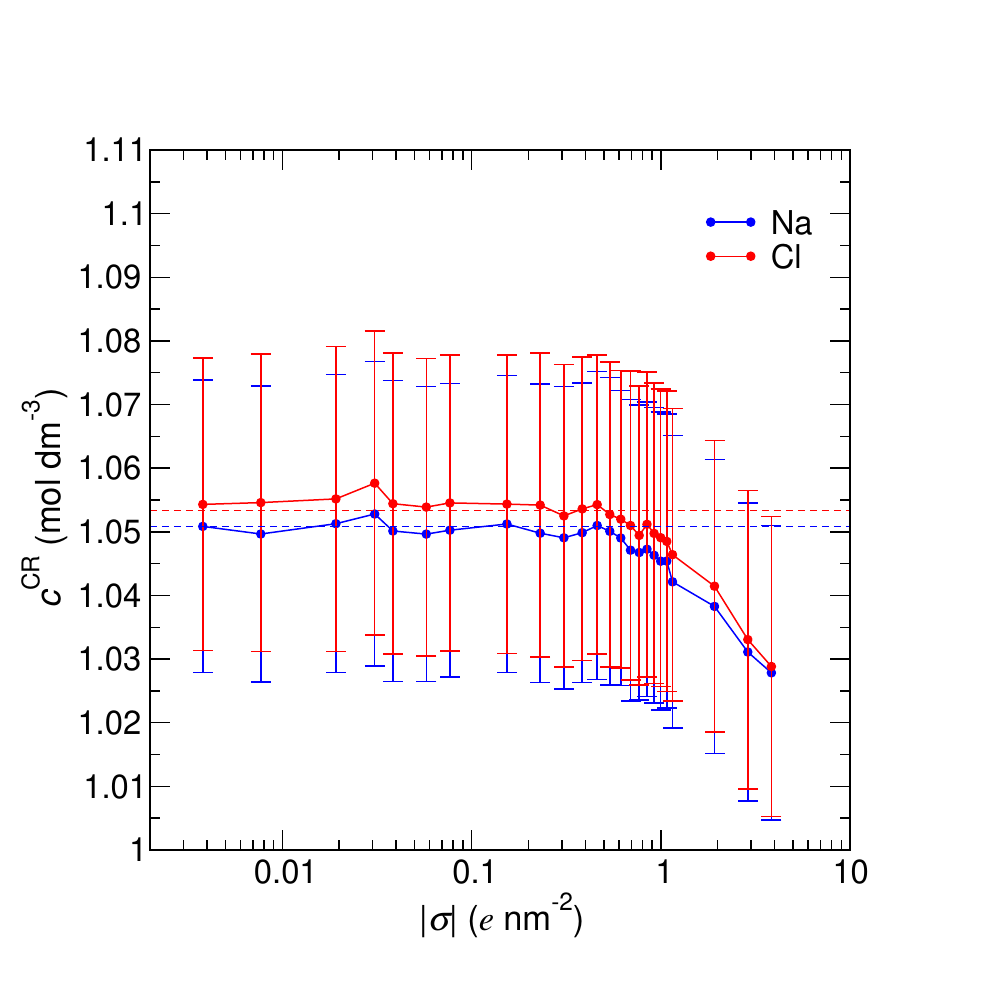}
  \caption[]{Mean concentrations of Na$^+$ and Cl$^-$ in the control regions ($c^{\mathrm{CR}}$) of C$\mu$MD simulations where the target NaCl concentration was 1~M and where surface charges, $\sigma$, were applied uniformly to the graphite surface. Error bars show the standard error in the mean values from 10~ns windows in 50~ns simulations and the dashed line provides the mean $c^{\mathrm{CR}}$ from simulations in the absence of graphite surface charges.}
  \label{fgr:electrode-CR-concn}
\end{figure}

The basal plane surface area was 29.75583~nm$^2$ and absolute values of charge $|q|=0.0001$--$0.1e$ were applied to the 1144 outermost carbon atoms.
This resulted in absolute surface charges densities,  $| \sigma |=0.0038$--$3.8 e~\mathrm{nm}^2$.
However, beyond $| \sigma | = $1 $e~\mathrm{nm}^2$, over-depletion of ions in the reservoir region occurred (see Figure \ref{fgr:electrode-CR-concn}). 
We therefore limited our analyses to systems where $| \sigma |$ was below $1 e~\mathrm{nm}^2$.
Table \ref{tab:charge} provides all of the surface charge densities analysed.
Electric fields from the applied surface charges were determined using the Poisson Equation (Equation \ref{eqn:poisson} in Section \ref{section:theory}).
Using a constant relative permittivity, $\varepsilon_r = 71$, to model the solvent medium, Grahame's Equation (Equation \ref{eqn:grahame} in Section \ref{section:theory}) predicts a potential drop across the interface, $\psi^0_{\mathrm{Grahame}}$, of 0.28--58~mV. 
As discussed in the main text, Grahame's Equation makes inaccurate assumptions about the double layer structure at intermediate--high concentrations; the potential change, therefore, is underestimated using this approach.

\begin{table}[ht]
\begin{center}
\begin{tabular}{ c|c|c } 
 $|q|$ ($e~\mathrm{atom}^{-1})$& $|\sigma|$ ($e$~nm$^2$) & E (V~nm$^{-1}$)  \\
\hline
0.0001 & 0.0038 & 0.07 \\
0.0002 & 0.0077 & 0.14 \\
0.0005 & 0.0192 & 0.35 \\
0.0008 & 0.0308 & 0.56 \\
0.0010 & 0.0384 & 0.70 \\
0.0015 & 0.0577 & 1.04 \\
0.0020 & 0.0769 & 1.39 \\
0.0040 & 0.1538 & 2.78 \\
0.0060 & 0.2307 & 4.17 \\
0.0080 & 0.3076 & 5.57 \\
0.0100 & 0.3845 & 6.96 \\
0.0120 & 0.4614 & 8.35 \\
0.0140 & 0.5382 & 9.74 \\
0.0160 & 0.6151 & 11.13 \\
0.0180 & 0.6920 & 12.52 \\
0.0200 & 0.7689 & 13.91 \\
0.0220 & 0.8458 & 15.31 \\
0.0240 & 0.9227 & 16.70 \\
0.0260 & 0.9996 & 18.09 \\
\end{tabular}
\caption{\label{tab:charge}Applied surface charges, $q$, and surface charge densities, $\sigma$, used in simulations in this work. Electric fields, E, calculated using the Poisson Equation (where $\varepsilon_r=1$) are also provided.}
\end{center}
\end{table}

\paragraph{Structural analyses using Plumed.}

\begin{figure}[ht]
\centering
  \includegraphics[width=0.4\linewidth]{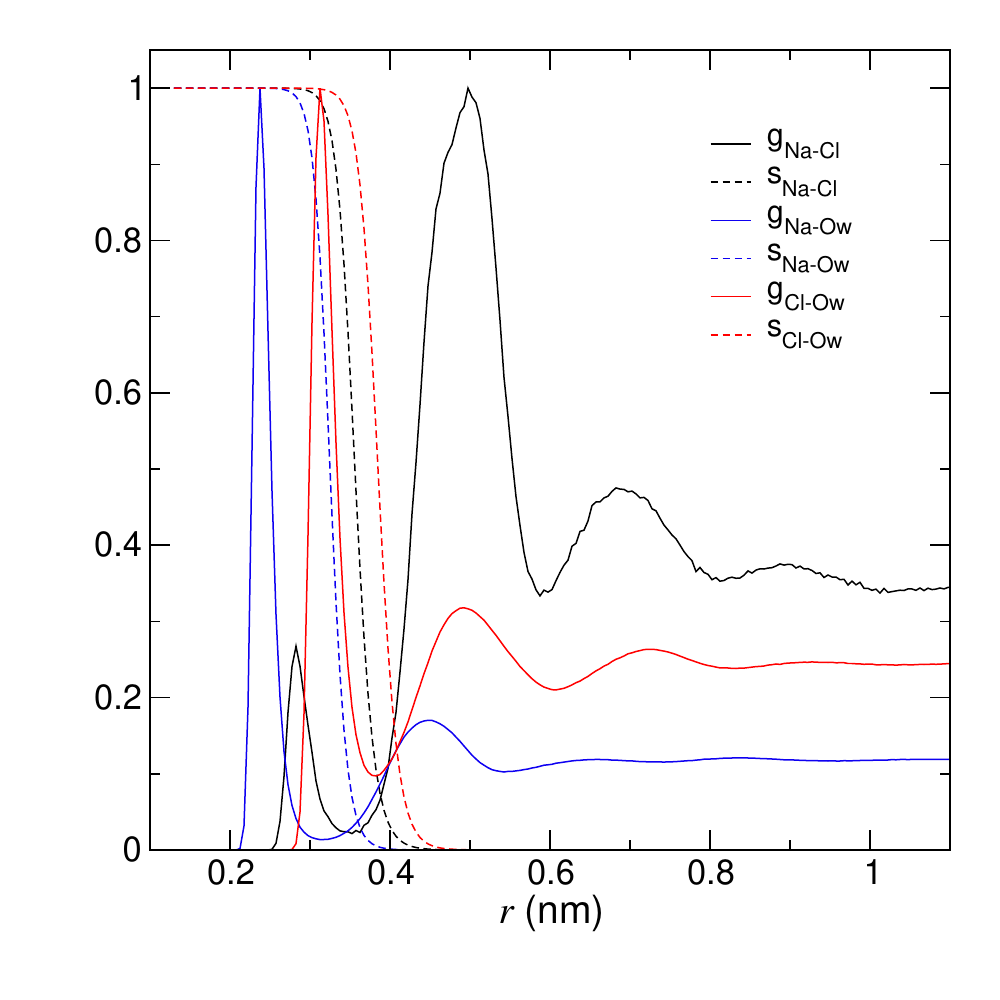}
  \caption[]{Pairwise radial distribution functions (RDFs), $g$, for atom types shown by the subscript labels in the figure key, calculated from a 10~ns bulk 1~M NaCl(aq) solution simulation. RDFs were normalised so that the maximum value is one. Also shown by the dashed lines are the switching functions, $s_{ij}$, used by the Plumed package to determine atoms coordinated in their respective first spheres. $s_{ij}=\frac{1-\left(\frac{r_{ij}}{r_0} \right)^{32}}{1-\left(\frac{r_{ij}}{r_0} \right)^{64}}$, where $r_{ij}$ are the distances between atoms $i$ and $j$, and with $r_0$ parameters provided in the main text.}
  \label{fgr:rdfs-s}
\end{figure}

Average first-sphere coordination numbers, $N_{i-j}$, between atoms $i$ and $j$ were calculated using a continuous but sharp definition of coordination implemented in Plumed: \citeSI{tribello1_plumed_2014}
\begin{equation}
N_{i-j}=\frac{1}{M_i}\sum_{i=1}^{M_i}\sum_{j=1}^{M_j}\frac{1-\left(\frac{r_{ij}}{r_0} \right)^{32}}{1-\left(\frac{r_{ij}}{r_0} \right)^{64}}
\end{equation}
where $M_i$ and $M_j$ are the numbers of atoms of types $i$ and $j$, respectively; $r_{ij}$ are the distances between ion pairs $i$ and $j$; and, $r_0$ were set to 0.355, 0.325 and 0.385~nm for the pairs Na--Cl, Na--Ow and Cl--Ow (where Ow is oxygen of water), respectively.
Figure \ref{fgr:rdfs-s} highlights that the chosen definitions of coordination ensured that only atoms in direct contact were considered when calculating the mean coordination numbers. 

To analyse ion clusters, we adopted a depth first search algorithm \citeSI{tribello1_analyzing_2017} to identify clusters from adjacency matrices built using the above definition of first-sphere coordination between ions. 
Clusters were defined using a continuous switching function to identify ion associates containing two or more ions. 
Instructions to access the Plumed input files for both coordination and cluster analyses are provided at the end of this section.

\paragraph{Diffusion coefficients.}

The diffusion coefficients, $D$, for ions and water were calculated using the Einstein relation: $D=\lim_{t\rightarrow{\infty}}d<(\mathbf{r}(t)-\mathbf{r}(0))^2>/6dt$, where $\mathbf{r}(t)$ is the position of a particle at time $t$. 
As a reference, $D$ for ions and water were calculated from a simulation of bulk NaCl(aq) at 1~M. 
In a simulation of NaCl(aq) containing 148 ions and 4000 water molecules in a cubic cell simulated for 10~ns at 298~K and 1~atm, $D=0.94 \pm 0.07, 1.26 \pm 0.08~\mathrm{and}~2.17 \pm 0.03~\times 10^{-5}~\mathrm{cm}^2~\mathrm{s}^{-1}$ for Na$^+$, Cl$^-$ and O of water, respectively. 
When the correction of Yeh and Hummer is applied to account for the finite size effects of simulationn cells,\citeSI{yeh_diffusion_2004} (using the shear viscosity of SPC/E water from Reference \citepSI{gonzalez_shear_2010}) the values of the corrected diffusion coefficients were $1.11, 1.44~\mathrm{and}~2.35~\times 10^{-5}~~\mathrm{cm}^2~\mathrm{s}^{-1}$.
This makes the calculated mean ion diffusion coefficients slightly smaller than those calculated by Joung and Cheatham but within the uncertainty estimates. \citeSI{joung_molecular_2009} 

Diffusion coefficients for ions and water were measured in $\Delta x=0.4$~nm windows moving away from the graphite surface in $50\times1$~ns calculations, with $D$ calculated by fitting to data when $t=100-600$~ps. 
This procedure was followed due to the propensity for ions and water molecules to translate to adjacent windows in $x$ on the timescales of the simulations. 
Mean $D$ were therefore estimated by averaging many shorter trajectories where the position of atoms at $t=0$ determined their contribution to mean $D$ values. 
We also calculated the $x$ component of the diffusion coefficients ($D_x$) to determine the mobility of molecular species when the electrical properties of the solution are changing in the double layer region. 

\paragraph{Additional files.}

GROMACS and Plumed input and example output files, including the force field parameters necessary to reproduce the simulation results reported in this paper, are available on github (see https://github.com/aaronrfinney/CmuMD-NaCl\_at\_graphite).
The PLUMED input files are also accessible via PLUMED-NEST (www.plumed-nest.org \citeSI{the_plumed_consortium_promoting_2019}), the public repository for the PLUMED consortium, using the project ID,  plumID:21.011. 
Details on how to use and implement the C$\mu$MD method within PLUMED is available on github (see https://github.com/mme-ucl/CmuMD). 

\subsection{Experimental Details}
\label{sec:expt}
The differential capacitance, $C^d$, of the NaCl(aq)--graphite system was measured via electrochemical impedance spectroscopy (EIS) using a droplet configuration. 
The top face of an HOPG crystal (1 cm$^2$, SPI-1 grade, Mosaic spread angle 0.4$^{\circ} \pm 0.1^{\circ}$; Structure Probe, Inc.) was exfoliated by peeling off a layer attached to adhesive tape (Scotch Tape, 3M). 
A 6 mm inner diameter silicone O-ring was placed on the freshly exfoliated face into which was pipetted a 200 \textmu{L} droplet of NaCl(aq) (99.999\% metal basis, Sigma Aldrich) solution (18.2 M$\Omega$~cm, Purelab Chorus, Elga). 
A reference electrode (leak-free Ag/AgCl, 3.4 M KCl, Innovative Instruments Ltd.) and a counter electrode (0.5 mm diameter Pt wire) were then quickly lowered into the droplet and the EIS was commenced. 
The impedance was measured at applied potentials between -0.5 and 0.0 V vs Ag/AgCl, using a 10 mV perturbation, at 105 Hz with a potentiostat (compactstat, Ivium Technologies B.V.).
Preliminary experiments showed that the impedance at this frequency was dominated by capacitance (the phase angle was 89$^\circ$) and was independent of the frequency over the entire concentration range studied. 
$C^d$ was therefore calculated directly from the imaginary part of the complex impedance, $Z"$, using $C^d = -(2\pi Z"fA)^{-1}$, where $f$ is the frequency of the perturbation and $A$ is the geometric area defined by the O ring. \citeSI{bard1_electrochemical_2001}

\begin{figure}[h]
\centering
  \includegraphics[width=0.5\linewidth]{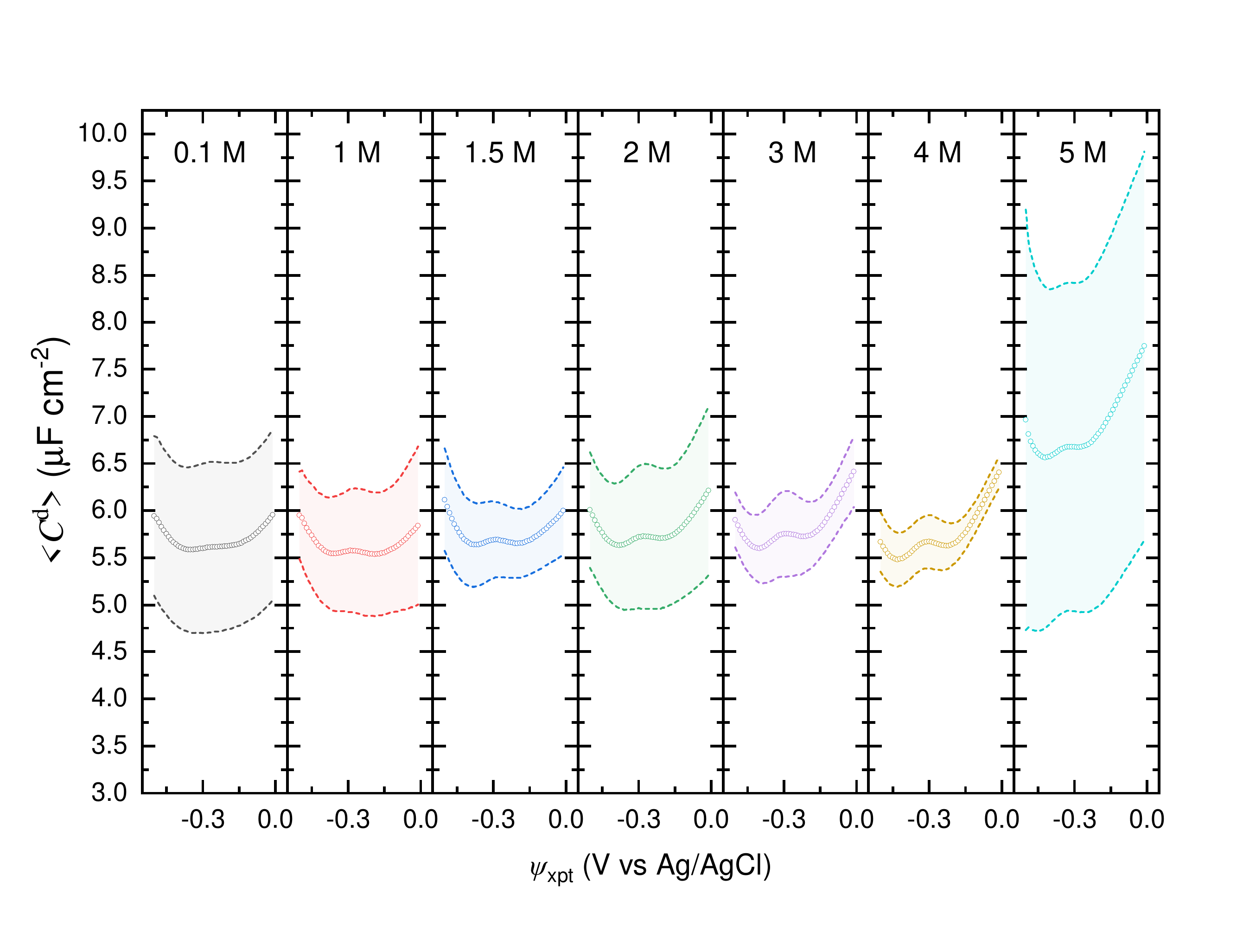}
  \caption[Experimental mean $C^d$-$\psi_{xpt}$ curves]{Experimental $C^d$-$\psi_{xpt}$ curves for HOPG in NaCl(aq) solutions of various concentration (indicated in the figure panels) measured using Protocol A (see SI Section \ref{sec:methods}). Plots show the mean (points) and standard deviation (dashed lines) of 3 measurements (or 2 measurements at 1 M and 3 M).}   
  \label{fgr:mean_Cd}
\end{figure}

Two protocols were adopted to measure the capacitance, with both emphasising the speed of measurement to preserve the integrity of the surface.  
Protocol A started from -0.5 V vs Ag/AgCl, with 5 s equilibration at this potential before the impedance was measured. 
The potential was then increased by 10 mV and the process repeated until the potential reached 0.0 V vs Ag/AgCl. 
In this way, the 51 potentials could be measured in several minutes, minimising the time that the HOPG was exposed to solution and, therefore, limiting the effect of potential organic impurity adsorption. 
Preliminary experiments, in which $C^d$ was measured as the potential was first incremented and then decremented, already showed a slight hysteresis indicative of surface deactivation.
Longer delays between measurements of 20 minutes showed substantial changes to both the magnitude and shape of the $C^d$--potential curve, in agreement with previous work,\citeSI{zou1_investigation_2016}  and confirming the need to proceed with measurements quickly.
Protocol B aimed to replicate the more common potential resolution used for measuring $C^d$--potential curves with 50 mV increments to the applied potential, initiated at -0.75 V vs Ag/AgCl. 
Protocol A was repeated at each sampled concentration, allowing uncertainties in $C^d$ (Figure \ref{fgr:mean_Cd}) to be estimated from the standard deviation of the measured data in three repeat experiments. 
Any curves that deviated from the common double minimum shape also showed much lower capacitance values and were therefore rejected. 

\begin{figure}[h]
\centering
  \includegraphics[width=0.5\linewidth]{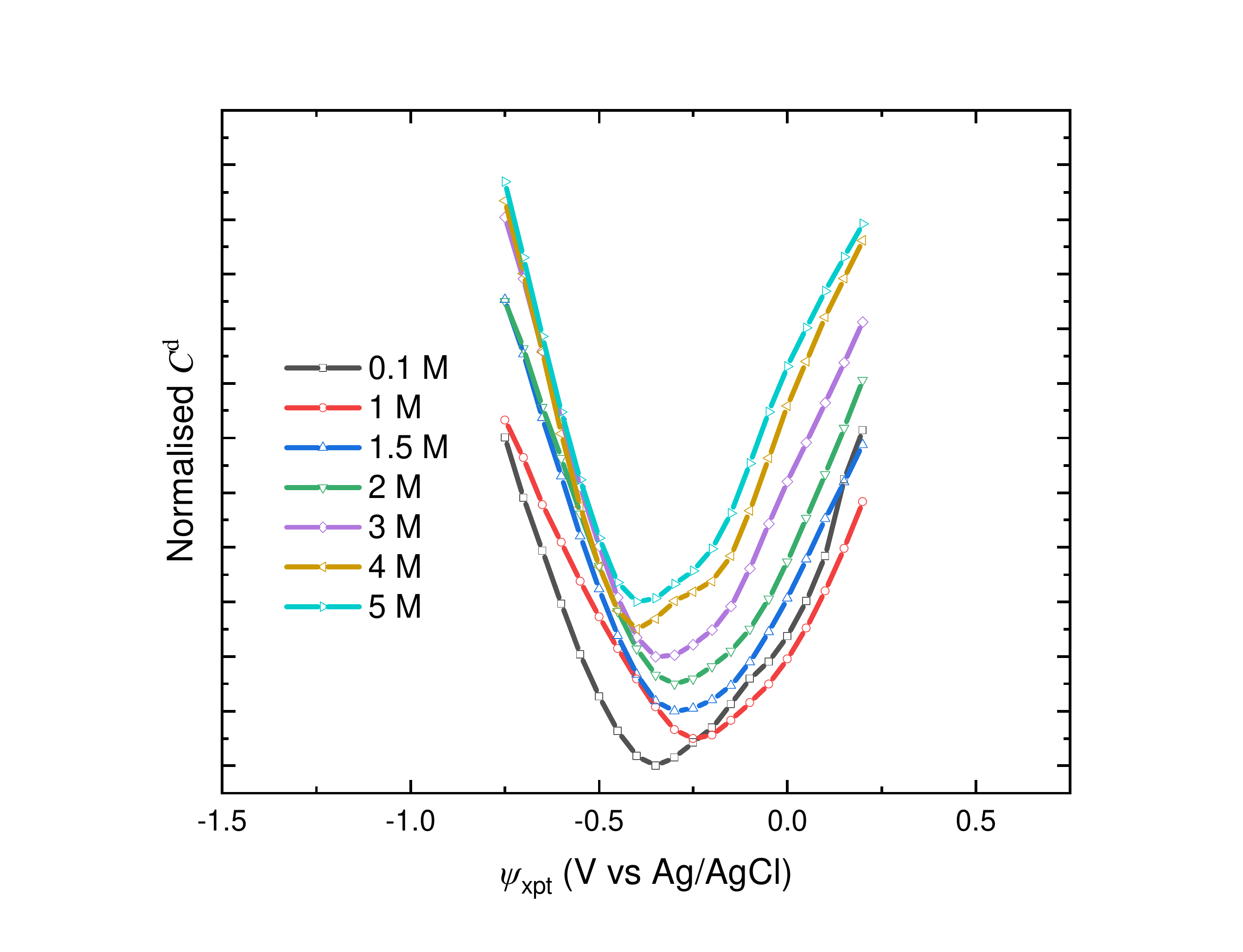}
  \caption[Experimental coarse $C^d$-$\psi_{xpt}$ curves]{Experimental $C^d$-$\psi_{xpt}$ curves for HOPG in NaCl(aq) solutions of various concentration (indicated in the plot) measured using Protocol B (see SI Section \ref{sec:methods}). For clarity $C^d$ has been normalised to the minimum value and the curves offset to show how the minimum shifts with changing bulk electrolyte concentrations.}   
  \label{fgr:coarse_Cd}
\end{figure}

Due to variations in the exposed surfaces formed during each exfoliation, the absolute magnitude of $C^d$ varies by 1--2~\textmu{F}~cm$^{-2}$ between measurements, as was seen previously,\citeSI{zou1_investigation_2016} with the value of $C^d$ at the minimum being the same at all concentrations (within error). 
In light of this result, only the shift in the $C^d$ minimum---taken to be the potential of zero charge (PZC)---was used for comparison with simulation results. The more negative of the two minima in $C^d$ was extracted for each repeat and used to calculate the mean and standard deviation of the $C^d$ minimum at each concentration. 
While it has been shown that asymmetric ion adsorption causes the PZC to shift away from the $C^d$ minimum, both change in the same direction and by similar amounts, and converge at high concentration, allowing min($C^d$) to function as an effective proxy for the PZC in our work.\citeSI{uematsu1_effects_2018} 
The shift in the $C^d$ minimum can also be seen in the data collected using Protocol B (see Figure \ref{fgr:coarse_Cd}), although the double minimum feature observed with Protocol A is not visible at this sampling resolution.

\section{Theoretical Background}
\label{section:theory}

Models to describe the double layer began with the work of Helmholtz in the 1870's.\citeSI{helmholtz_studien_1879} 
Since then, focus has largely been devoted to the effect of charged electrodes in contact with electrolyte solutions. 
Helmholtz suggested that a static, compensating layer of ions from the extended liquid phase adsorbs at a surface to neutralise the total surface charge. 
This so-called electrical condenser predicts a linear change in the electric potential, $\psi$, across a distance determined by the radius of the adsorbed charge carriers. 
The Helmholtz model fails to account for the thermal motion of ions at the interface. 

Gouy and Chapman provided the earliest theory able to account for the entropy of charge carriers in
solution adjacent to a charged planar surface.\citeSI{gouy1_sur_1910,chapman1_li_1913} 
Gouy-Chapman theory predicts that a diffuse ion layer assembles at the interface with a solid substrate. 
The surface charge density, $\sigma$, is thus compensated by the total charge density, $\rho$, in the liquid phase in the direction, $x$, orthogonal to the surface:
\begin{equation}
    \sigma = - \int_0^{\infty} \rho(x) \; dx 
    %\sigma = - \int_0^{\infty} \rho(x) \; dx = - \varepsilon \; \mathrm{E}_{el}
\label{eqn:sigma}
\end{equation}
Ions in the diffuse layer dissipate the electric potential of the double layer by adopting a structure normal to the surface consistent with a Boltzmann distribution:
\begin{equation}
    n_i(x) = n_i^{\mathrm{b}} \; \mathrm{exp} \left( -\beta z_i e \psi(x)\right) 
    \label{eqn:boltzmann}
\end{equation}
where $n_i$ is the ion $i$ number density and $n_i^{\mathrm{b}}$ is the uniform ion number density in the extended liquid phase (i.e. the bulk of the solution); $z_ie=q_i$, which is the electric charge of $i$ with $z_i$ valency and $e$ is the elementary unit of charge; and, $\beta=1/k_{\mathrm{B}}T$ where $k_{\mathrm{B}}$ is Boltzmann's constant and $T$ is temperature. 
The above equation highlights that the work required to bring ion $i$ from the extended liquid phase to the interfacial region, $W_i(x)= -z_i e \psi(x)$; hence, a fundamental assumption is that only electrical work is involved. \citeSI{facci_useful_2014}

The dependence of ion density distributions on ion charges implies a local departure from electroneutrality within the diffuse layer. 
The resulting $\psi$ and electric field, $\mathrm{E}$, in the double layer can be calculated using the Poisson equation from the charge densities and permittivity of the medium, $\varepsilon=\varepsilon_0\varepsilon_r$ (where $\varepsilon_0$ and $\varepsilon_r$ refer to the permittivity of a vacuum and the relative permittivity of the solution medium, respectively):

\begin{equation}
    \frac{d^2\psi(x)}{dx^2} = \frac{d\mathrm{E}(x)}{dx} = \frac{-\rho(x)}{\varepsilon }
\label{eqn:poisson}
\end{equation}
noting that $\rho=\sum_i z_ien_i$. 
Combining equations \ref{eqn:boltzmann} and \ref{eqn:poisson} leads to the non-linear Poisson-Boltzmann equation (PBE) for a binary univalent electrolyte ($|z_{\mathrm{cation}}|=|z_{\mathrm{anion}}|=1$ and $n^{\mathrm{b}}=n^{\mathrm{b}}(\mathrm{cation})=n^{\mathrm{b}}(\mathrm{anion})$):
\begin{equation}
    \frac{d^2\psi(x)}{dx^2} = \frac{en^{\mathrm{b}}}{\varepsilon} \left[ \mathrm{exp} \left( \beta e \psi(x) \right) - \mathrm{exp} \left( -\beta e \psi(x) \right) \right] 
\label{eqn:pb}
\end{equation}
This self-consistent, second order differential equation directly relates the charge distribution of ions to the potential of mean force associated with bringing a point charged ion into the diffuse layer (due to Coulombic intermolecular interactions). 
In typical applications of the PBE, mean field theory applies: local ion--ion and ion--solvent correlations are neglected and the electric potential is due to the thermally averaged electric field. 
Indeed, in this description, the solvent is implicit and is often modelled assuming a fixed value of $\varepsilon$. 
At the inner- and outer-most boundaries of the double layer, the following conditions apply, $\mathrm{E}^0=-\sigma\varepsilon^{-1}$ and $\mathrm{E}^\infty=0$.

To obtain the electrolyte densities at the electrode surface ($n_i^{\mathrm{s}}$), the derivative of Equation \ref{eqn:boltzmann} with respect to $x$ can be written as, 
\begin{equation}
\begin{split}
%    \color{blue} \frac{d n_i}{dx} &\color{blue} = - \beta z_i e n_i^0 \left( \frac{d \psi(x)}{dx} \right) \mathrm{exp} (- \beta z_i e \psi(x)) \\
%    \color{blue} \mathrm{as,} \; & \color{blue} \frac{d^2\psi(x)}{dx^2} \varepsilon = z_ien_i^0 \mathrm{exp} \left( -\beta z_i e \psi(x) \right) \\
%    \color{blue} \frac{d n_i}{dx} &\color{blue} = \beta \varepsilon \left( \frac{d^2 \psi(x)}{dx^2} \right) \left( \frac{d \psi(x)}{dx} \right) \\
    \frac{d n_i}{dx} &= \frac{\varepsilon \beta}{2} \frac{d}{dx} \left( \frac{d \psi(x)}{dx} \right)^2 
\label{eqn:concderiv}
\end{split}
\end{equation}
by adopting the relationship in Equation \ref{eqn:pb} for a generic ion type. \citeSI{facci_useful_2014} Integrating Equation \ref{eqn:concderiv} from the extended liquid phase to any $x$ position in the diffuse layer results in,
\begin{equation}
     \sum_i n_i(x) = \sum_i n_i^{\mathrm{b}} + \frac{\varepsilon \beta}{2} \left( \frac{d \psi(x)}{dx} \right)_x^2 
\label{eqn:ciz}
\end{equation}
Making use of the relationships in Equations \ref{eqn:poisson} and \ref{eqn:sigma}, Equation \ref{eqn:ciz} can be rewritten as,
\begin{equation}
     \sum_i n_i^{\mathrm{s}} = \sum_i n_i^{\mathrm{b}} + \frac{\sigma^2 \beta}{2 \varepsilon}
\label{eqn:ciel}
\end{equation}

Equation \ref{eqn:boltzmann}, and Equation \ref{eqn:pb} by association, indicate a diffuse layer of infinite size: $c_{\mathrm{cation}}=c_{\mathrm{anion}}$ when $\psi(x\rightarrow{\infty})=0$.
Yet, the extent to which $n_i$ decay and converge to $n^{\mathrm{b}}$ within thermal fluctuations can be approximately determined. 
The size of the diffuse layer can be calculated according to the Gouy length, $\lambda_{\mathrm{G}}$, which is determined as a solution to Equation \ref{eqn:pb}:

\begin{equation}
     \lambda_{\mathrm{G}} = \frac{2 \kappa^{-2}}{\mathrm{exp}(\beta e \psi(x)/2)+\mathrm{exp}(-\beta e \psi(x)/2)}
\label{eqn:lg}
\end{equation}

\noindent Here, $\kappa^{-1}$ is the Debye length which takes the functional form,
\begin{equation}
    \kappa^{-1}=\left( \frac{\varepsilon}{2\beta e^2 n^{\mathrm{b}}} \right)^{\frac{1}{2}}
\label{eqn:debye}
\end{equation}
The Gouy length decreases exponentially as $\sigma$ is increased; however, at the highest surface charge densities, this can be approximated as $2\varepsilon/\beta e \sigma$. \citeSI{gray_nonlinear_2018} 
As $\sigma$ approaches zero, $\lambda_{\mathrm{G}}$ and $\kappa^{-1}$ converge (at constant $n^{\mathrm{b}}$). 
Indeed, for surfaces with a low applied potential---where $e|\psi|<k_{\mathrm{B}}T$---the Debye length is assumed to be the characteristic decay length over which ion concentrations converge to $n^{\mathrm{b}}$. 
The size of the double layer, therefore, decreases as $n^{\mathrm{b}}$ increases. 
When dealing with moderate electrolyte concentrations or surface charge densities, these simple models offer limited predictive power.
A general solution to the Poisson-Boltzmann equation can be provided for ideal solutions at perfectly planar, homogeneously charged surfaces.\citeSI{butt_physics_2003} 
The added complexity of non-ideal electrolyte solutions and imperfect surface geometries makes predicting the size of the diffuse layer a considerable challenge.

To relate the surface charge density to the surface potential, one can employ Equation \ref{eqn:sigma} with Equation \ref{eqn:pb} rewritten using a hyperbolic trigonometric function,
leading to a notable result known as Grahame's equation\citeSI{grahame1_electrical_1947}: 
\begin{equation}
\begin{split}
%    \color{blue} \frac{d^2\psi(x)}{dx^2} &\color{blue} = \frac{en^0}{\varepsilon} \left[ \mathrm{exp} \left( \beta e \psi(x) \right) - \mathrm{exp} \left( -\beta e \psi(x) \right) \right] \\
%    & \color{blue} = \kappa^2 \mathrm{sinh} \left( \beta e \psi(x) \right) \\
%    \color{blue}  \frac{d\psi(x)}{dx} & \color{blue} = -2 \kappa \mathrm{sinh} \left( \frac{\beta e \psi(x)}{2} \right) \\
    \sigma &= - \int_0^{\infty} \rho(x) \; dx = - \varepsilon \; \frac{d\psi(x)}{dx} \bigg|_{x=0} \\
    &= 2 \varepsilon \kappa \mathrm{sinh} \left( \frac{\beta e \psi(0)}{2} \right) \\
    &= \sqrt{8 \varepsilon n^{\mathrm{b}} \beta^{-1}}  \mathrm{sinh} \left( \frac{\beta e \psi^0}{2} \right)
\label{eqn:grahame}
\end{split}
\end{equation}

The Stern modification to the Gouy-Chapman model accounts for the finite size of ions in the immediate vicinity of the surface. 
In the Gouy-Chapman-Stern (GCS) model, a layer of ions adsorbed at the substrate/solution interface is introduced as a fixed plate capacitor.\citeSI{stern1_zur_1924} 
The ions in such layers are generally considered to be immobile.
Following this approach, the double layer can be divided into two regions: an innermost region called the Stern layer, where the potential decays linearly, followed by a diffuse layer, where the potential decays exponentially according to the PBE. 
The potential at the transition point between these two regions is associated with the $\zeta$-potential. 
Given an assumed discontinuity in the mobility of charge carriers at this point, the characteristic distance from the electrode interface is described as the slipping plane. 
The depth of the Stern layer is of the size of ion radii, with or without their solvation sphere(s), labelled as the inner- and outer-Helmholtz planes, respectively. 

Despite their practical relevance, the simple mean-field models so far described fail to account for the asymmetric adsorption of cations and anions (with equal but opposite applied surface potential), and the role of solvent molecules---of particular importance in the case of highly polar solvents---when determining the electric potential. 
Moreover, it follows from Equation \ref{eqn:ciel} that, in the absence of surface charges, the net total ion concentration at the interface is unperturbed compared to that in the extended liquid phase. 
This is inconsistent with the fact that uncharged surfaces can, and do, induce perturbations to the structure and composition of the liquid due to van der Waals forces and steric effects leading to an effective surface charge.\citeSI{uematsu1_effects_2018,attard1_ion_1995} 
The potential of zero charge, $\psi^{\mathrm{pzc}}$, encapsulates these phenomena and can be calculated using Equation \ref{eqn:poisson} with zero applied surface potential. 
In systems containing metallic electrodes, a linear correlation in $\psi^{\mathrm{pzc}}$ with respect to the work function of the electrode is often observed.\citeSI{schmickler_electronic_1996}
According to Gouy-Chapman theory, $C_{\mathrm{d}}$ has a minimum at $\psi^{\mathrm{pzc}}$; however, specific ion adsorption leads to deviations depending upon the ionic species in contact with the electrode.\citeSI{uematsu1_effects_2018} 
At high electrolyte concentrations, asymmetric capacitance--potential curves accompany a shift in the minimum capacitance, when cation and anion surface-adsorption energies differ.\citeSI{uematsu1_effects_2018} 
At relatively high ionic strengths, electrolytes in the double layer can also influence the charge distribution of the substrate. 
This polarisation affects the value of $C_{\mathrm{d}}^{\mathrm{Stern}}$.\citeSI{zhan1_computational_2017} 

Extensions to the simplest mean field models and alternative models to predict the electrical properties of interfaces have been proposed.\citeSI{uematsu1_effects_2018,fedorov1_ionic_2014,dogonadze_solvation_1988,conway_modern_1979} 
Particularly relevant in this regard is the case of ionic liquids at electrified interfaces, where the anisotropy and asymmetry in the adsorption of charge carriers in the liquid phase cannot be ignored. 
Models which e.g., involve minimising the free energy function describing the system constructed using a Poisson--Boltzmann lattice-gas model, have been developed to describe these systems.\citeSI{kornyshev1_double-layer_2007,fedorov1_ionic_2014}. 
It was noted that these models are not so valuable to understand conventional electrolyte systems---such as the one that is the focus of this study---which should be modelled reasonably well by Gouy-Chapman theory.\citeSI{kornyshev1_double-layer_2007}

\section{C$\mu$MD Bulk Concentrations}
\label{sec:concentration}

Some discrepancies were observed between the targeted ion molar concentrations ($c^t$) and those that were measured in the control regions ($c^{\mathrm{CR}}$) of C$\mu$MD simulations following equilibration. 
In particular, there was a slight contraction of the targeted parameter space, with the measured $c^{\mathrm{CR}}$ being greater and smaller than $c^t$ at the lower and higher end of the overall targeted concentration range. 
The deviations were most apparent when ion concentrations in the reservoir were much lower/higher than $c^{\mathrm{CR}}$ (see Table \ref{tab:conc}). 
Large concentration gradients between the CR and reservoir also induced fluctuations in $c(x)$ around $x_F$. 
While not important in the current work, these effects can be mitigated by varying the values of $k$ and $\omega$ in the definition of $F_i^{\mu}$, and by ensuring that the reservoir concentration does not become too low. 

\begin{figure}[ht]
\centering
  \includegraphics[scale=1]{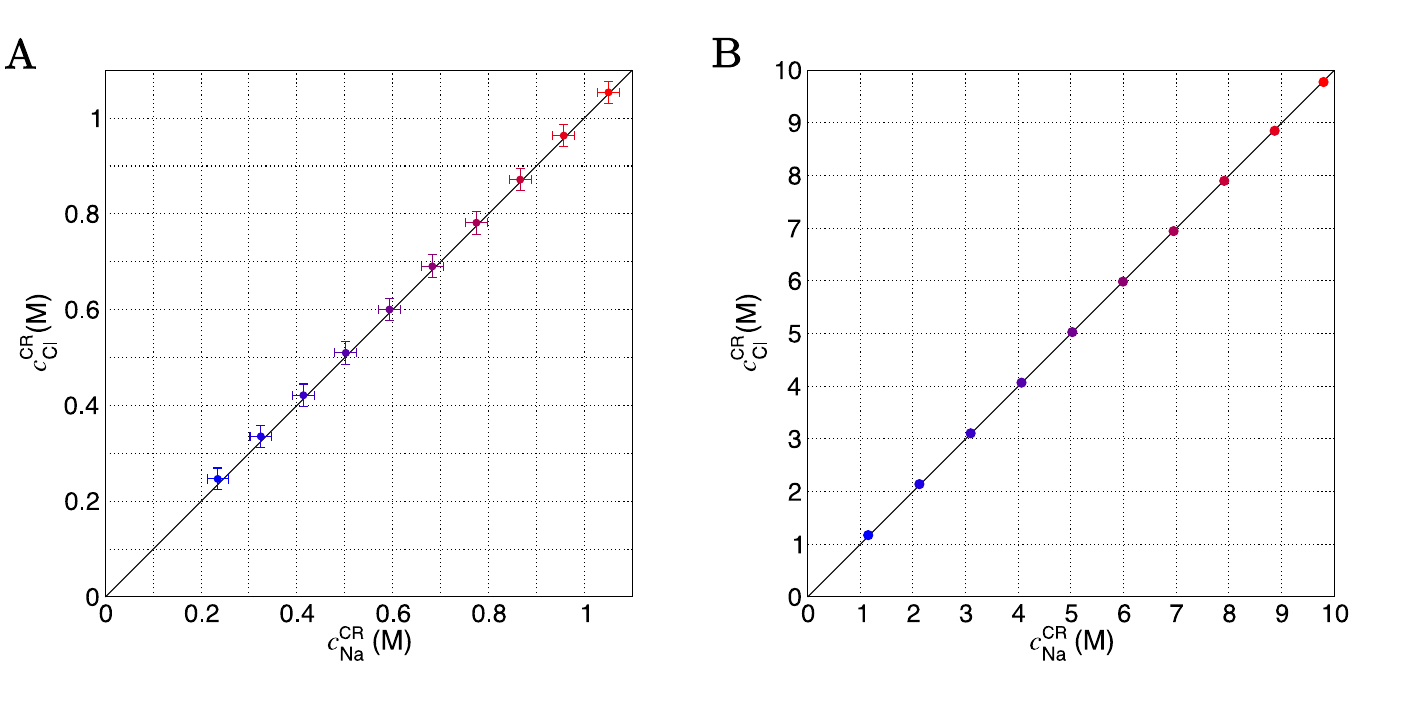}
  \caption[]{Na$^+$ and Cl$^-$ ion concentrations measured in the control regions of C$\mu$MD simulations ($c^{\mathrm{CR}}$). A provides data from simulations where target concentrations, $c^t=$ 0.1--1 M in 0.1  M increments. B provides data from simulations where $c^t=$ 1--10 M in 1 M increments. The concentration range is highlighted by the blue$\rightarrow{\mathrm{red}}$ colour scale indicating increasing $c^t$. Error bars (smaller than the data points in B) highlight uncertainties of one standard deviation in the data gathered from the final 50~ns of simulation.}
  \label{fgr:cr-concentrations}
\end{figure}

In all our analyses, we take $c^{\mathrm{b}}$ to be the mean NaCl concentration in a 0.5~nm stable region of the $c(x)$ profile in the CR (i.e., the mean value of $c(x)$ in the CR away from $x_F$).
$c^{\mathrm{b}}$ is therefore the `bulk' concentration of ions---representing the extended solution phase---in equilibrium with the solution at the graphite surface. 
Importantly, fluctuations in $c^{\mathrm{CR}}$ and $c^{\mathrm{b}}$ were independent of $c^t$, as shown in Table \ref{tab:conc} and Figure \ref{fgr:cr-concentrations}. 
This is essential e.g., to determine the depth of the double layer region and for the accurate calculation of ion activities as a function of $x$.

\begin{table}[th]
\begin{center}
\begin{tabular}{ c|c|c|c } 
 & $c^t$ (M) & $c^{\mathrm{b}}$ (M) & $c^{\mathrm{b}}/c^t$ \\
\hline
\multirow{10}{5em}{System A} & 0.1 & 0.23(0.01)& 2.3 \\ 
&  0.2 & 0.32(0.01)& 1.62 \\
&  0.3 & 0.41(0.01)& 1.38 \\
&  0.4 & 0.49(0.02)& 1.22 \\
&  0.5 & 0.58(0.02)& 1.17 \\
&  0.6 & 0.67(0.02)& 1.12 \\
&  0.7 & 0.78(0.02)& 1.11 \\
&  0.8 & 0.87(0.02)& 1.08 \\
&  0.9 & 0.95(0.03)& 1.06 \\
&  1.0 & 1.05(0.03)& 1.05 \\ 
\hline
\multirow{10}{5em}{System B} & 1.0 & 1.20(0.03) & 1.2 \\ 
&  2.0 & 2.20(0.03) & 1.1 \\
&  3.0 & 3.20(0.04) & 1.07 \\
&  4.0 & 4.11(0.05) & 1.03 \\
&  5.0 & 5.01(0.05) & 1 \\
&  6.0 & 5.83(0.05) & 0.97 \\
&  7.0 & 6.67(0.07) & 0.95 \\
&  8.0 & 7.53(0.08) & 0.94 \\
&  9.0 & 8.39(0.11) & 0.93 \\
& 10.0 & 9.23(0.07) & 0.92 \\
\end{tabular}
\caption{\label{tab:conc}Target mean NaCl concentrations, $c^t$, and measured bulk NaCl concentrations, $c^{\mathrm{b}}$, in the CR regions of C$\mu$MD simulations. Uncertainties, shown in parentheses, indicate one standard deviation in the mean concentration values (taken from the final 50~ns of simulation trajectories) measured in a 0.5~nm region of the CR away from $x_F$.}
\end{center}
\end{table}

\section{Double Layer Size from Solution Mass Densities}
\label{section:dlsize}
The size of the interfacial solution region can be identified by calculating the position in $x$ where the solution mass density, $\rho_m=\sum_i m_i n_i$ (where $m_i$ is the mass of atom $i$), diverges from the bulk uniform mass density. 
The first derivative of $\rho_m$ with respect to $x$ was calculated after first applying a Savitzky-Golay filter to smooth the data. 
The size of the interface region in solution was then determined as the distance where the fluctuations in $d \rho_m /dx$ in a 0.5~nm region moving away from the graphite surface were within $2 \sigma_b$; here, $\sigma_b$ is the standard deviation of $d \rho_m /dx$ in the bulk region. 

\begin{figure}[ht]
\centering
  \includegraphics[scale=0.7]{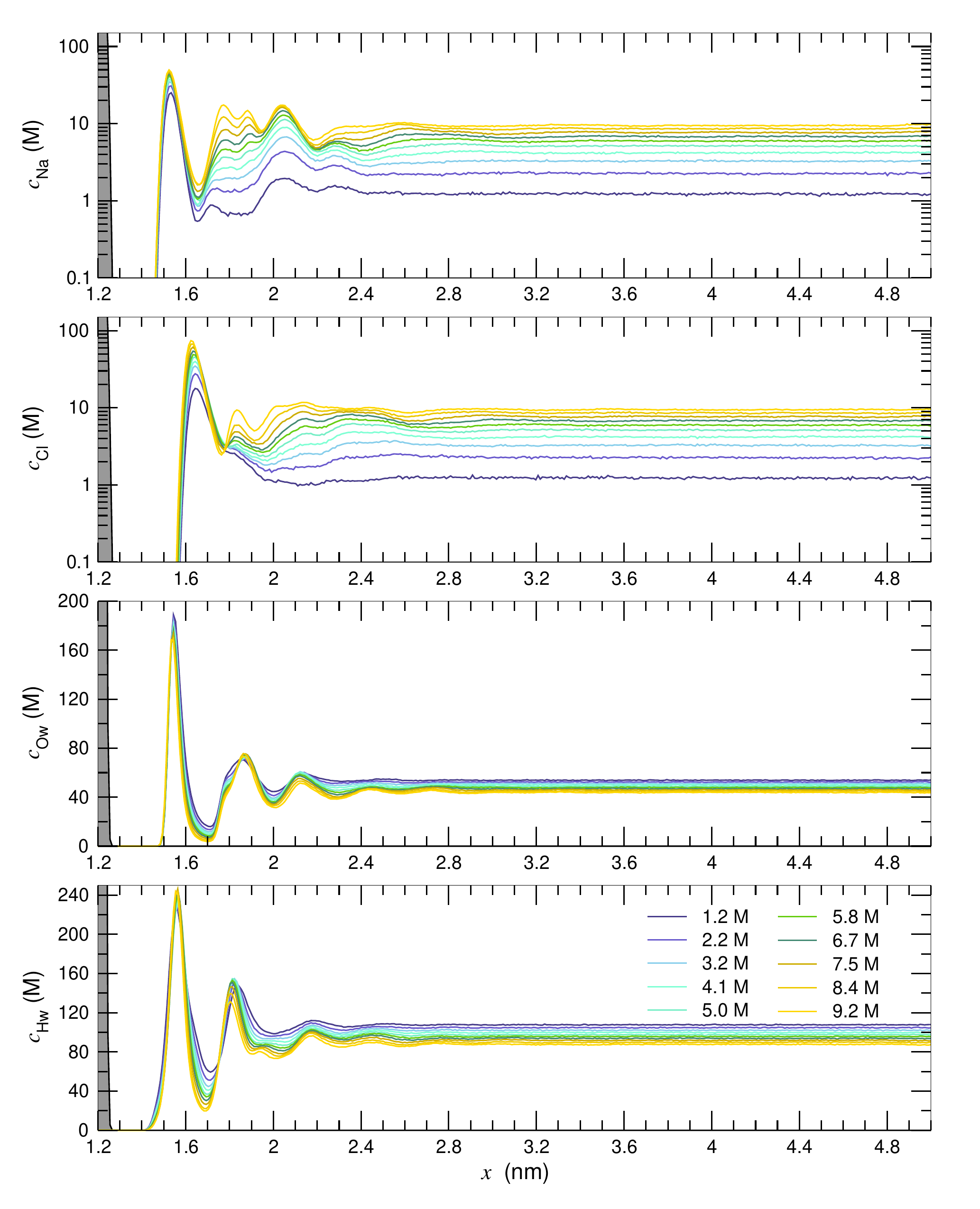}
  \caption[]{Averaged atom molar concentrations, $c$, measured for Na$^+$, Cl$^-$, Ow (water oxygen) and Hw (water hydrogen) atoms in C$\mu$MD simulations of System B (the upper end of the total concentration range; see SI Section \ref{sec:methods}) as a function of distance, $x$, from the midpoint in the simulation cell. The colour scale indicates the mean molarity of ions in the bulk region as indicated by the key in the bottom panel. Ion concentrations are provided using a logarithmic scale. The edge of the carbon surface is highlighted by the grey peak on the left of the $x$ axis.}
  \label{fgr:rho-profiles2}
\end{figure}

The double layer size determined following the density criterion above is approximately constant (within statistical uncertainties) as a function of bulk ion concentrations. 
The mean value of the double layer size was $1.43 \pm0.25$~nm. The result is not surprising, given that $\rho_m$ is dominated by the density of water. 
The interface induces ordering of the water molecules in a relatively compact region near the graphite surface, but this does not extend beyond five molecular layers across the entire concentration range.
An exception to this was the highest concentration case ($c_{\mathrm{NaCl}}^{\mathrm{b}}=$9.8 M), where the width of the density perturbation induced by the graphite was $2.43 \pm 0.47$~nm.
We note that at the highest bulk ion concentrations, changes to the ordering of water molecules within the double layer region emerges, as indicated by the concentration profiels in Figure \ref{fgr:rho-profiles2}.

\clearpage

\newpage
\section{Additional Figures}

\begin{figure}[ht]
\centering
  \includegraphics[scale=0.7]{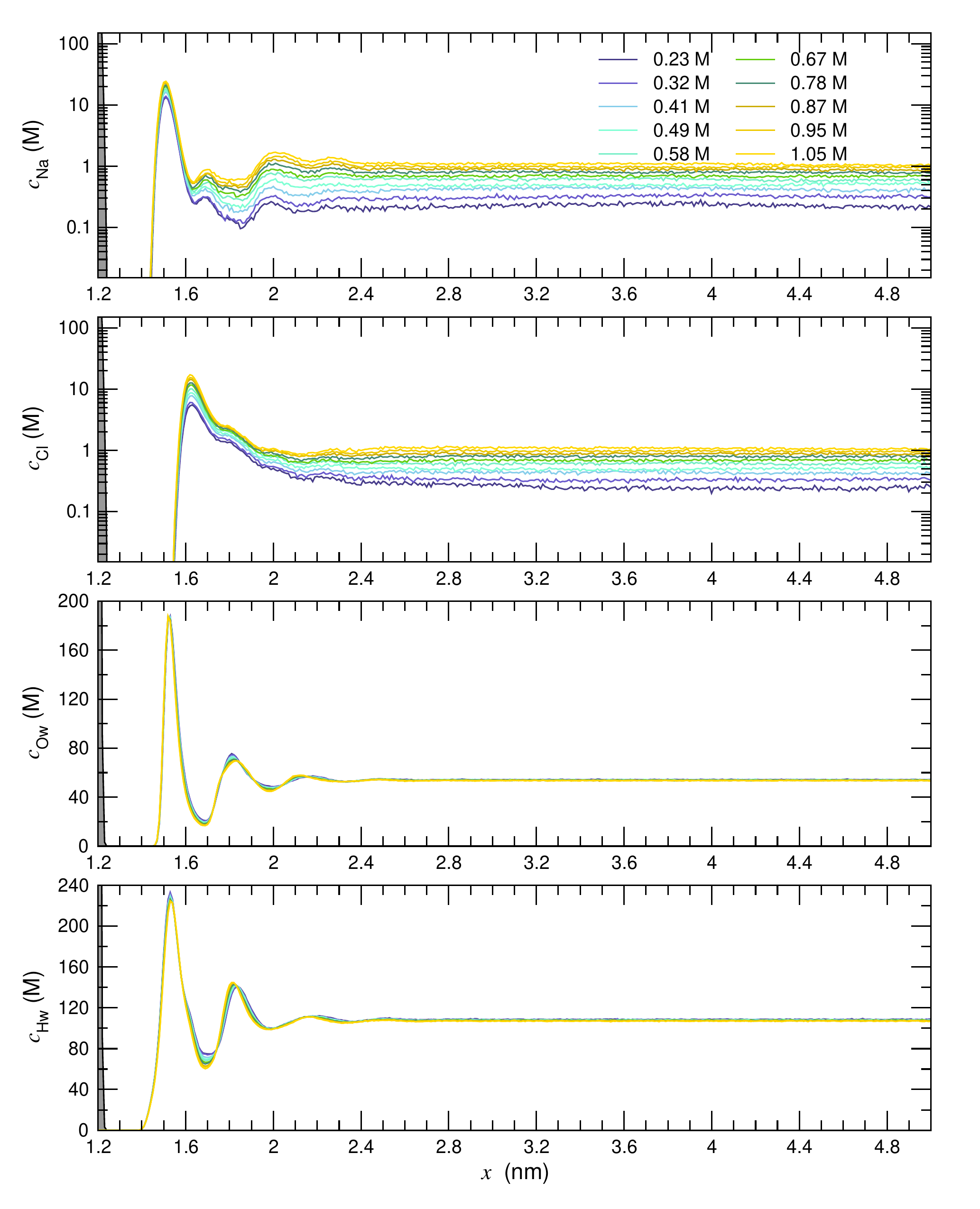}
  \caption[]{Averaged atom molar concentrations, $c$, measured for Na$^+$, Cl$^-$, Ow (water oxygen) and Hw (water hydrogen) atoms in C$\mu$MD simulations of System A (the lower end of the total concentration range; see SI Section \ref{sec:methods}) as a function of distance, $x$, from the midpoint in the simulation cell. The colour scale indicates the mean molarity of ions in the bulk region as indicated by the key in the top panel. Ion concentrations are provided using a logarithmic scale. The edge of the carbon surface is highlighted by the grey peak on the left of the $x$ axis.}
  \label{fgr:rho-profiles1}
\end{figure}

\begin{figure}[ht]
\centering
  \includegraphics[scale=0.8]{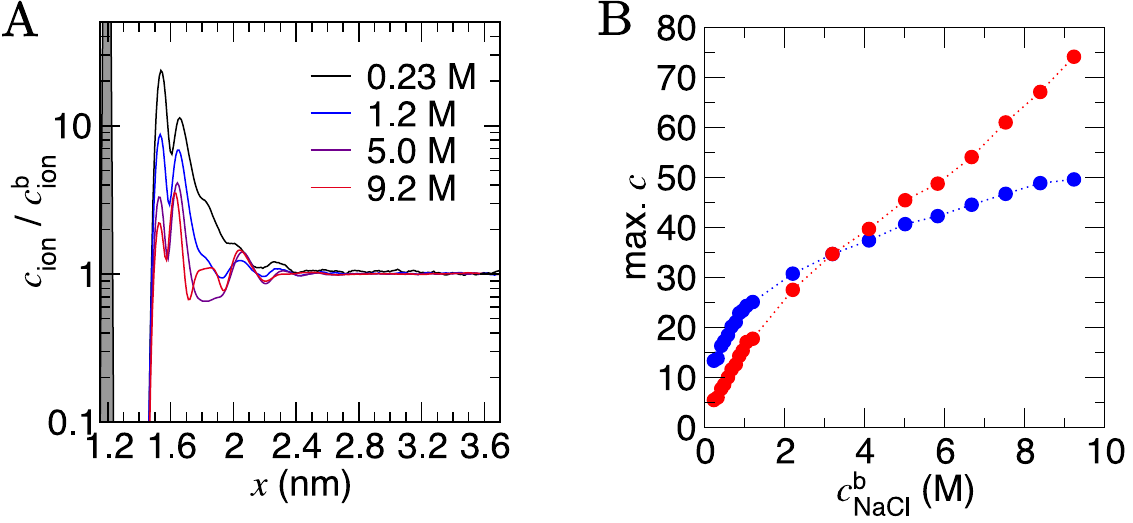}
  \caption[]{A: Total ion concentrations ($c_{\mathrm{ion}}=c_{\mathrm{Na}}+c_{\mathrm{Cl}}$) as a function of $x$, normalised by concentrations in the bulk ($c^{\mathrm{b}}_{\mathrm{ion}}$) as indicated by the key.
  B: Maximum Na$^+$ (blue) and Cl$^-$ (red) concentrations in the interfacial region taken from the concentration profiles in Figures \ref{fgr:rho-profiles1} and \ref{fgr:rho-profiles2}}. 
  \label{fgr:max-concn}
\end{figure}

\begin{figure}[ht]
\centering
  \includegraphics[scale=0.75]{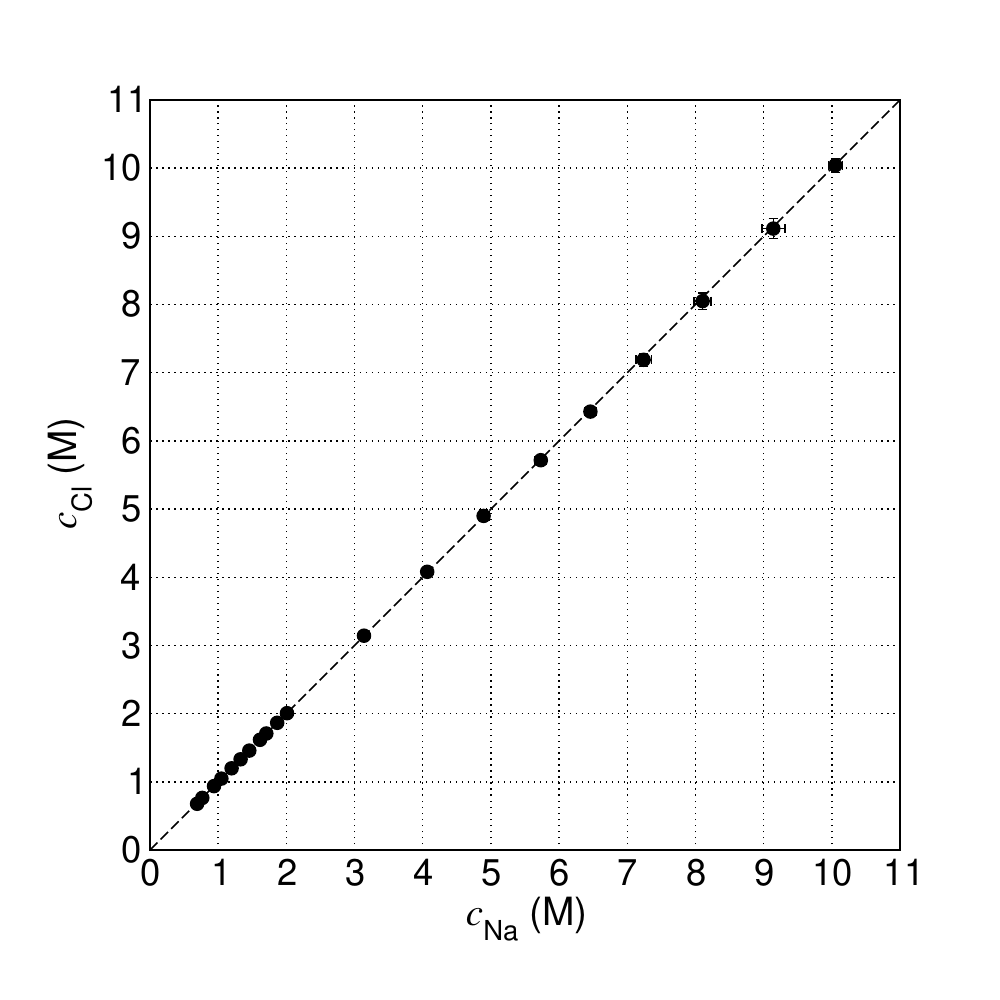}
  \caption[]{Na$^+$ and Cl$^-$ ion concentrations measured in the interface regions of C$\mu$MD simulations ($c^{\mathrm{i}}$). The interface was identified as a 1.5 nm region from the distance of closest approach of ions to the graphite surface in the solution phase. Error bars indicate uncertainties of one standard deviation in the data gathered from the final 50~ns of simulation.}
  \label{fgr:int-concn}
\end{figure}

\begin{figure}[ht]
\centering
  \includegraphics[width=0.85\linewidth]{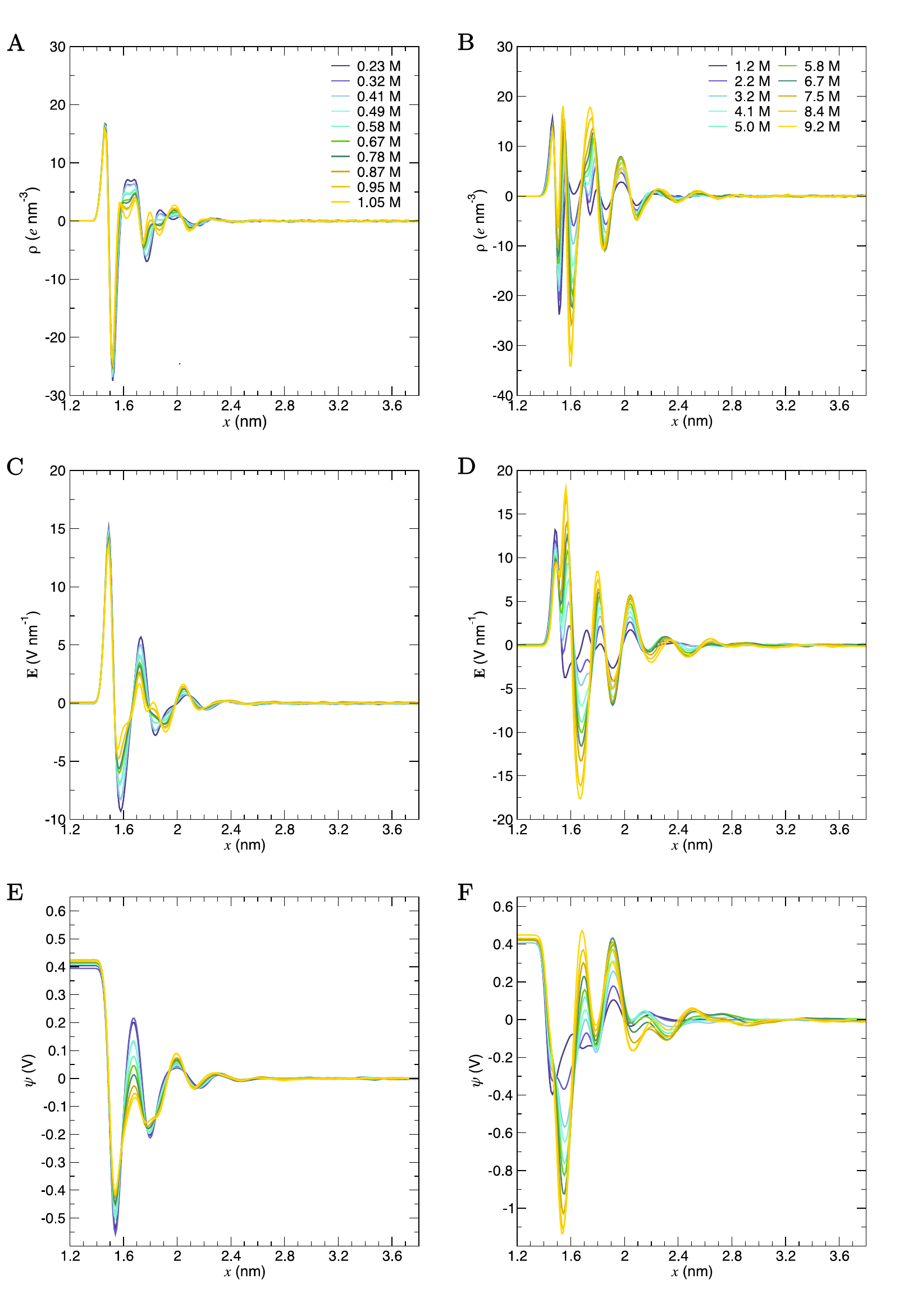}
  \caption[]{Solution charge densities, $\rho$ (A and B), electric fields, E (C and D) and electric potential, $\psi$ (E and F), as a function of distance from the centre of the simulation cell, $x$, in C$\mu$MD simulations with varying bulk NaCl concentrations shown by the legends in A (for A, C and D) and B (for B, D and F).}
  \label{fgr:poisson-plots}
\end{figure}

\begin{figure}[ht]
\centering
  \includegraphics[width=0.45\linewidth]{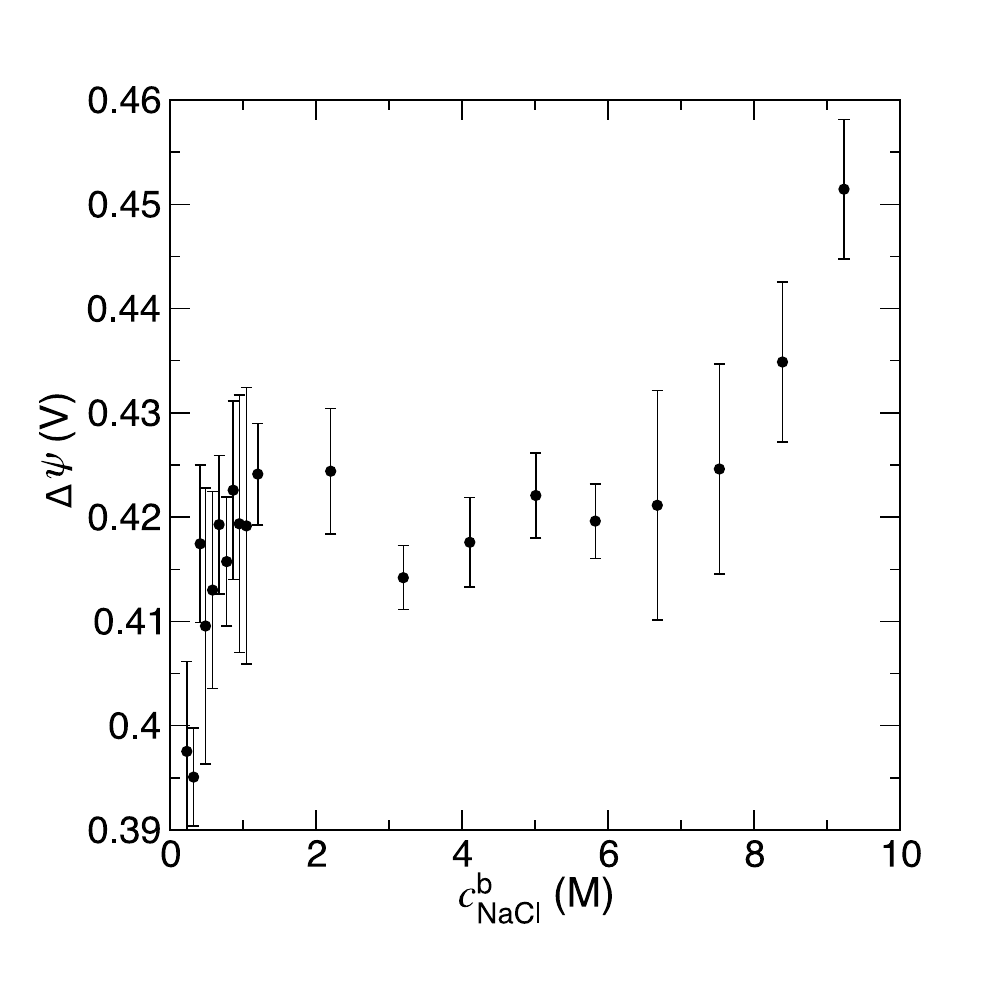}
  \caption[]{The electric potential change across the double layer, $\Delta \psi$, as a function of $c^{\mathrm{b}}_{\mathrm{NaCl}}$ when all solution species are included in the calculation. Error bars highlight the standard error of the mean taken from 10 ns windows over the final 50 ns of simulation.}
  \label{fgr:poisson-pzc}
\end{figure}

\begin{figure}[ht]
\centering
  \includegraphics[width=0.4\linewidth]{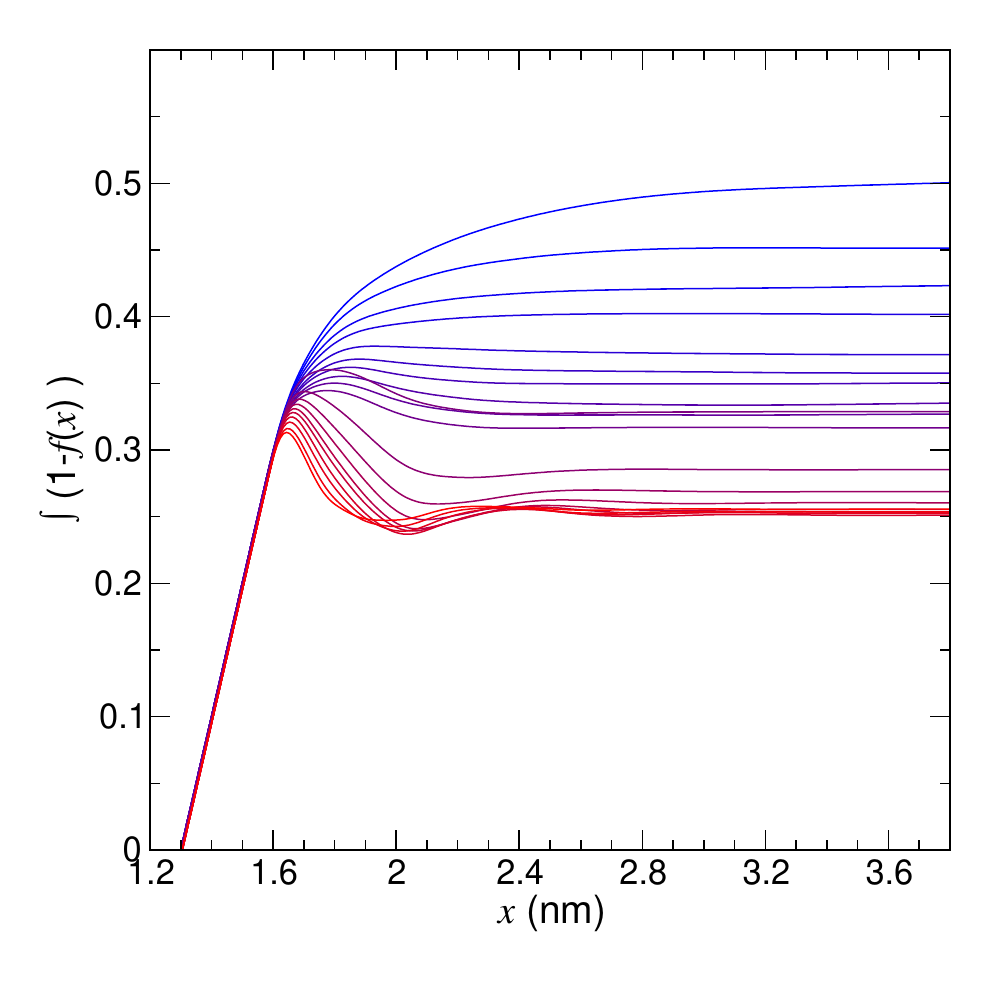}
  \caption[]{The cumulative integral of $(1-f)$---with $f$ being the screening factor presented in Figure \ref{fgr:pzc-info} of the main paper---as a function of $x$. The colour scale from blue to red indicates increasing bulk ion concentrations, $c_{\mathrm{NaCl}}^{\mathrm{b}}$, over the entire sampled concentration range.}
  \label{fgr:intscreening}
\end{figure}

\begin{figure}[ht]
\centering
  \includegraphics[width=0.6\linewidth]{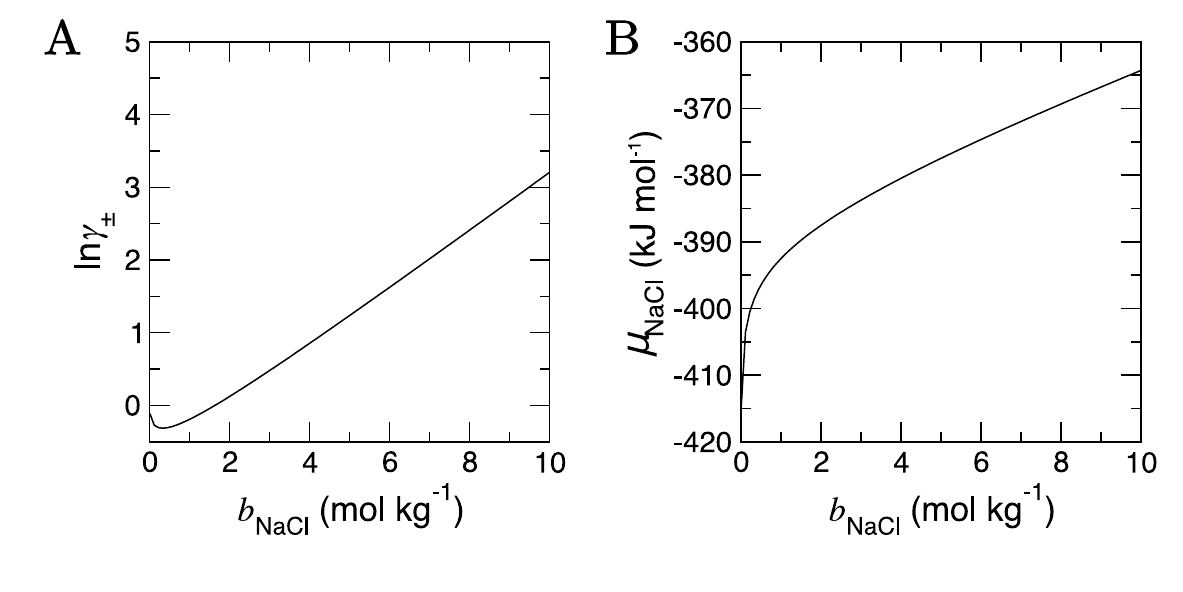}
  \caption[]{Mean ion activity coefficient, $\gamma_\pm$, and chemical potential of ions in solution, $\mu_{\mathrm{NaCl}}$, calculated for a range of electrolyte molalities, $b_{\mathrm{NaCl}}$, using the model Equations \ref{eq:chempot-model} and \ref{eq:activity-coeff} in the main paper.} 
  \label{fgr:chempotmodel}
\end{figure}

\begin{figure}[ht]
\centering
  \includegraphics[width=0.4\linewidth]{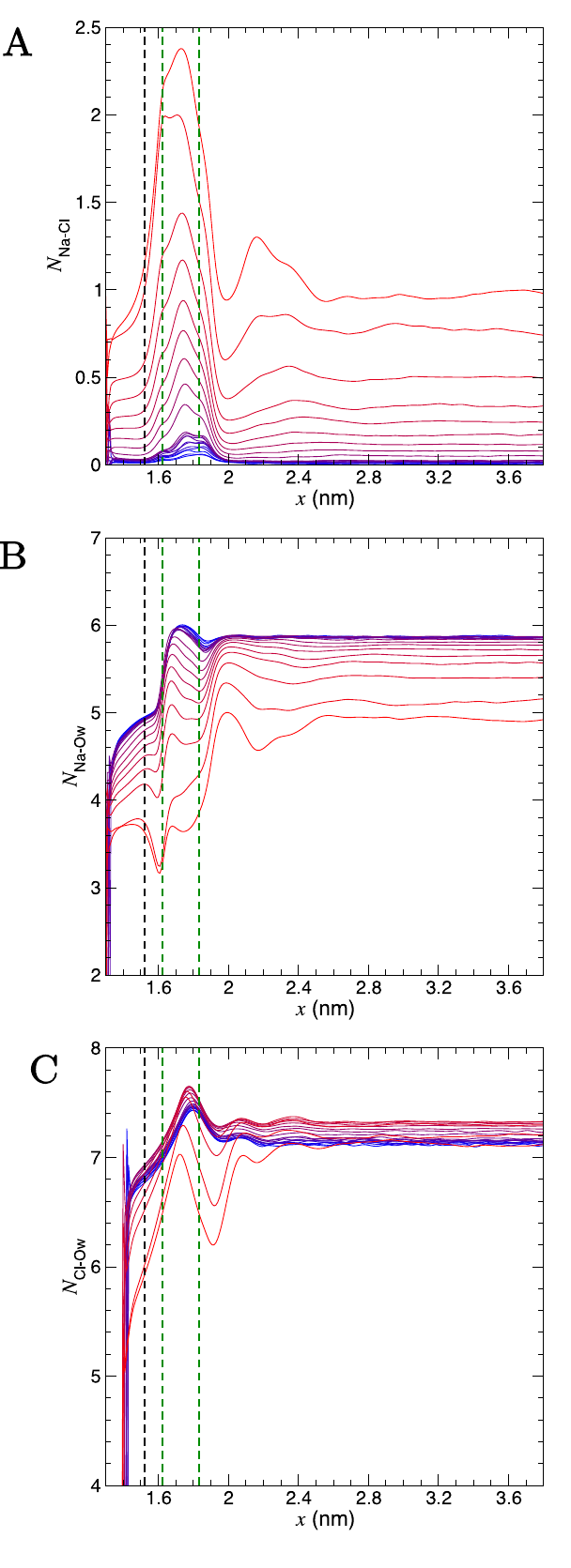}
  \caption[]{Average coordination numbers, $N$, as a function of $x$ for Na$^+$ with Cl$^-$ (A), Na$^+$ with O of water molecules (Ow; B) and Cl$^-$ with O of water molecules (Ow; C). The colour scale from blue to red indicates increasing bulk ion concentrations, $c_{\mathrm{NaCl}}^{\mathrm{b}}$, across the entire concentration range sampled. The black and green dashed lines indicate the maximum first Na$^+$ and first two Cl$^-$ densities in the concentration profiles for Na$^+$ and Cl$^-$ highlighted in Figure \ref{fgr:density-profiles} of the main paper. Note that the large fluctuations apparent where the distribution goes to zero are due to large statistical uncertainties in regions where the atom number densities also approach zero.}
  \label{fgr:CNprofiles}
\end{figure}

\begin{figure}[ht]
\centering
  \includegraphics[width=0.75\linewidth]{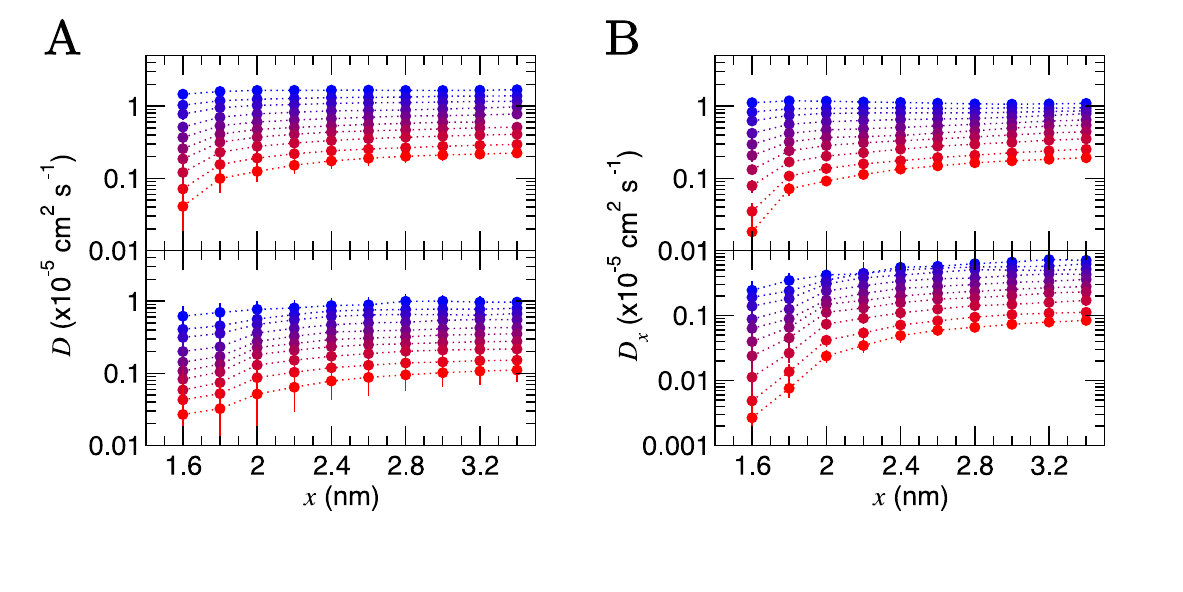}
  \caption[]{A: Diffusion coefficients, $D$, calculated in 0.4~nm regions of $x$ moving away from the graphite surface. $D$ is calculated from C$\mu$MD simulations with varying bulk concentrations of ions increasing from 1.2 to 9.2~M as shown by the colour scale from  blue to red. B provides $D_x$: the $x$ component of $D$. Error bars indicate uncertainties in the mean values of $D$ and $D_x$ from $50 \times 1$~ns trajectory windows. Top and bottom panels in A and B provide diffusion coefficients for water and ions, respectively.} 
  \label{fgr:diffusion-all}
\end{figure}

\begin{figure}[ht]
\centering
  \includegraphics[width=0.7\linewidth]{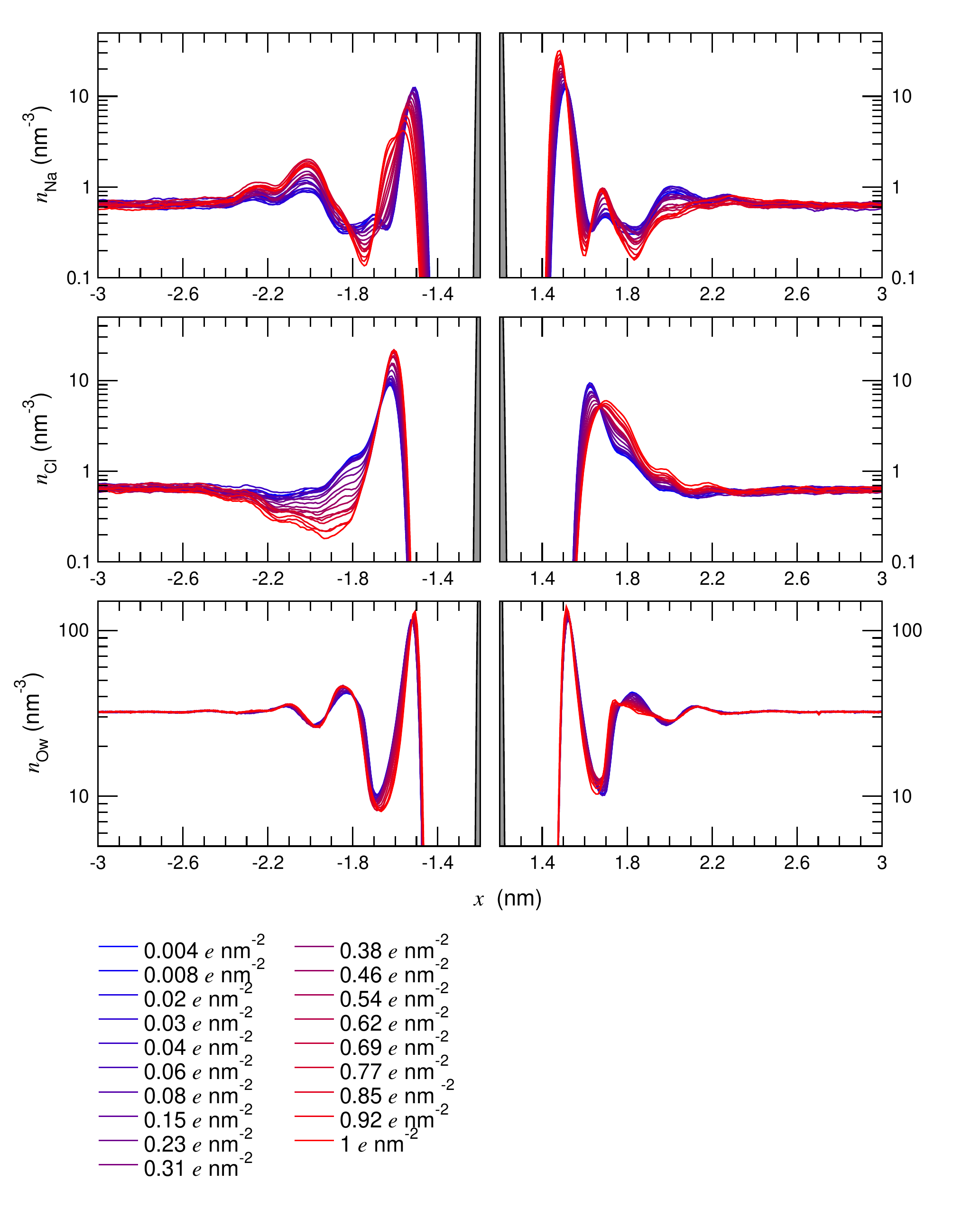}
  \caption[]{Number densities ($n$) of atom types, shown by the subscript labels (with Ow indicating water oxygen), as a function of $x$ in C$\mu$MD simulations. Charges were applied to the graphite surface where absolute charge densities, $\sigma$, increased from 0.004 to 1 $e$~nm$^{-2}$, as indicated by the blue $\rightarrow{}$red colour scale and explicitly listed in the key. Left panels are the densities on a logarithmic scale, where positively charged surfaces were exposed to solution, and right panels show the corresponding solution densities at the negatively charged face of the graphite slab. The grey peaks show the position of the edge of the graphite basal surface.}
  \label{fgr:electrode-atomdensities}
\end{figure}

\begin{figure}[ht]
\centering
  \includegraphics[width=0.7\linewidth]{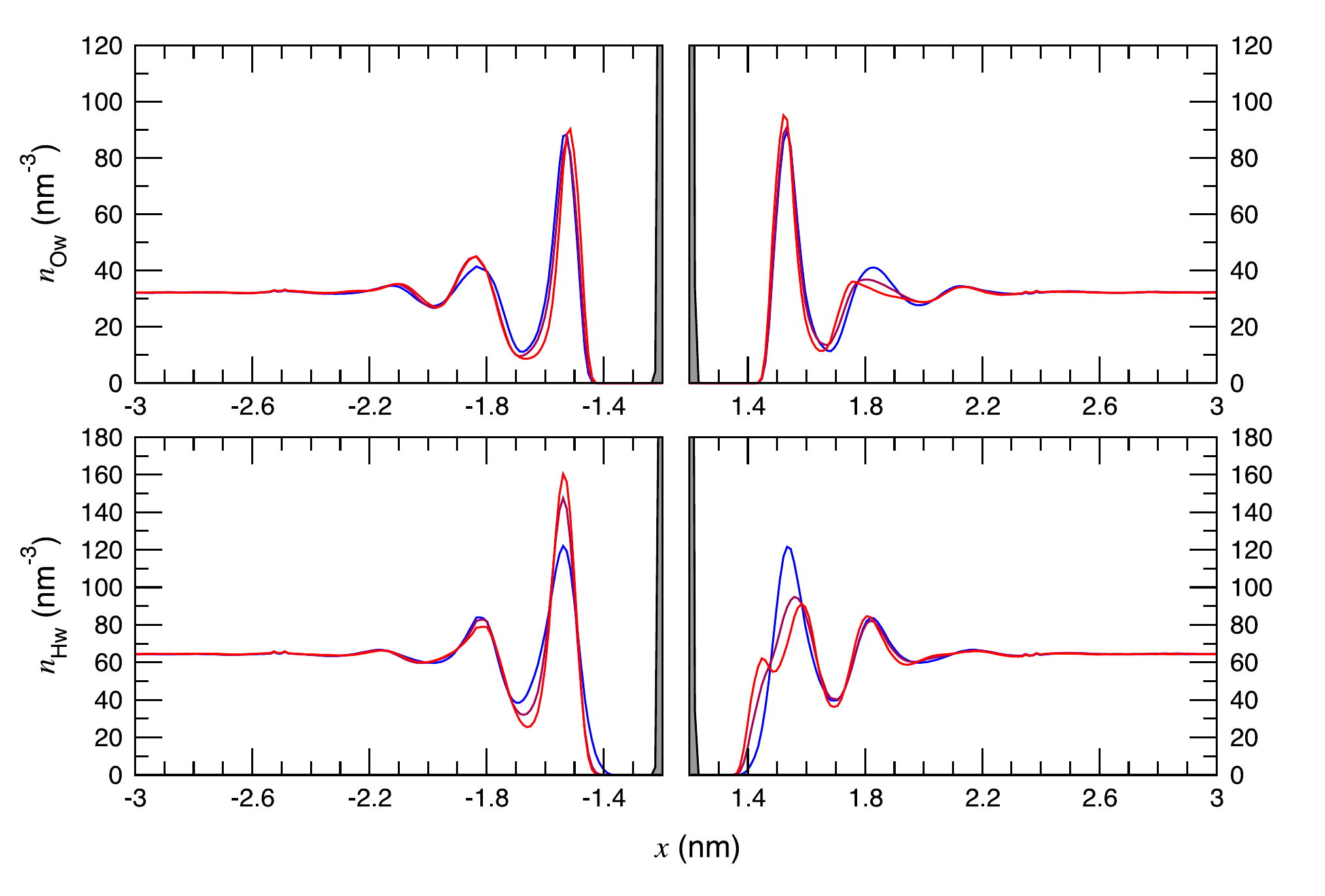}
  \caption[]{Water atom number densities ($n$) shown by the subscript labels Ow and Hw indicating water oxygen and hydrogen, respectively, from C$\mu$MD simulations with charges applied to the graphite surface. Surface charge densities, $\sigma$, were 0.004, 0.54 and 1 $e$~nm$^{-2}$, indicated by the use of blue, maroon and red lines, respectively. Left panels are the densities where positively charged surfaces were exposed to solution, and right panels show the corresponding number densities at the negatively charged face of the graphite surface. The grey peaks show the position of the edge of the graphite basal slab.}
  \label{fgr:electrode-waterdens-comp}
\end{figure}

\begin{figure}[ht]
\centering
  \includegraphics[width=0.7\linewidth]{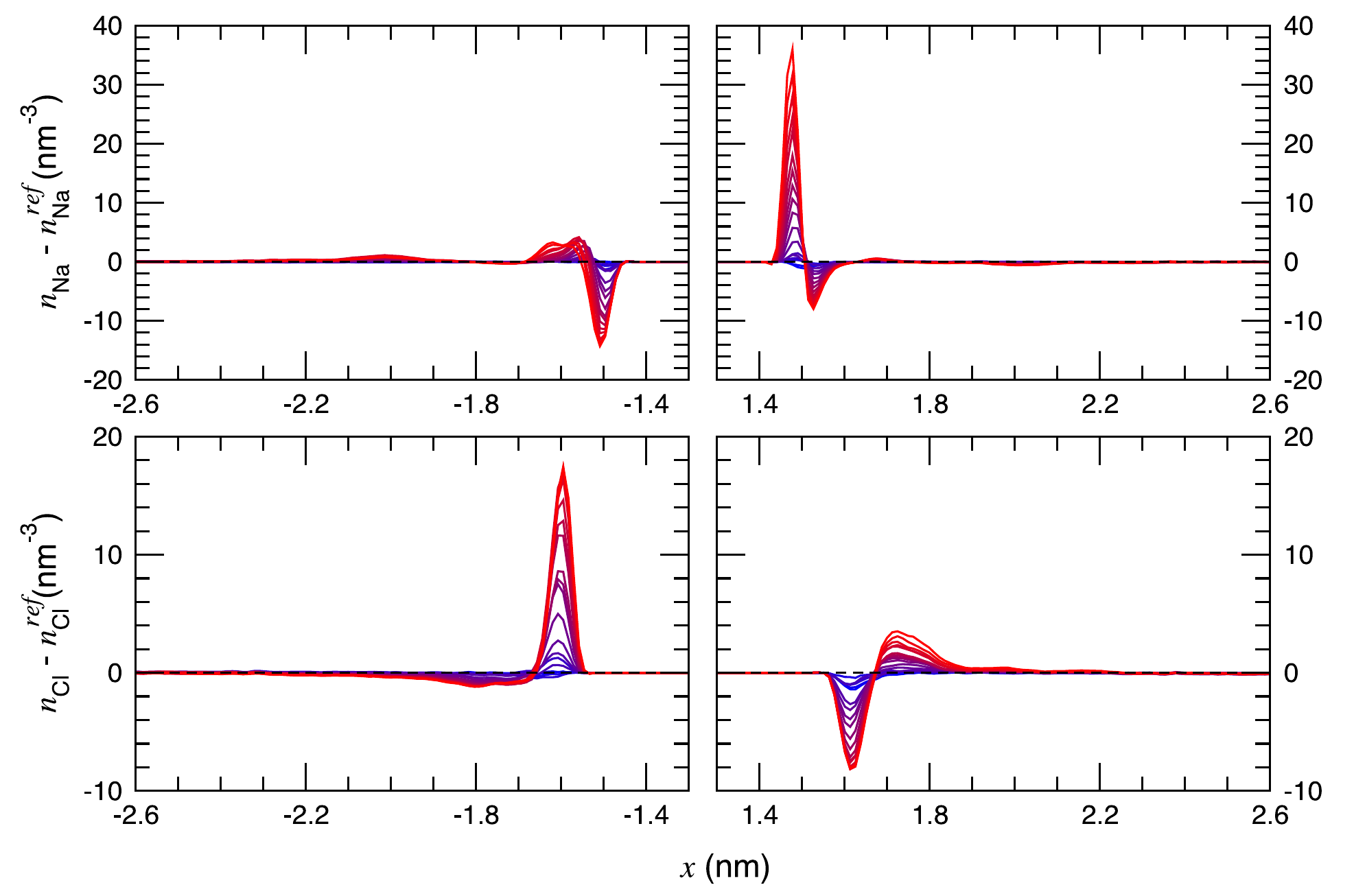}
  \caption[]{Excess ion number densities were calculated according to, $n-n^{ref}$, where $n$ and $n^{ref}$ are the ion number densities with and without applied graphite surface charges. $n-n^{ref}$ profiles in $x$ are provided for different values of surface charge (0.004--1 $e$~nm$^{-2}$) indicated by the blue $\rightarrow{}$red colour scale, and the surface charge densities are explicitly listed in the key in Figure \ref{fgr:electrode-atomdensities}.}
  \label{fgr:electrode-residual-densities}
\end{figure}

\begin{figure}[ht]
\centering
  \includegraphics[width=1.0\linewidth]{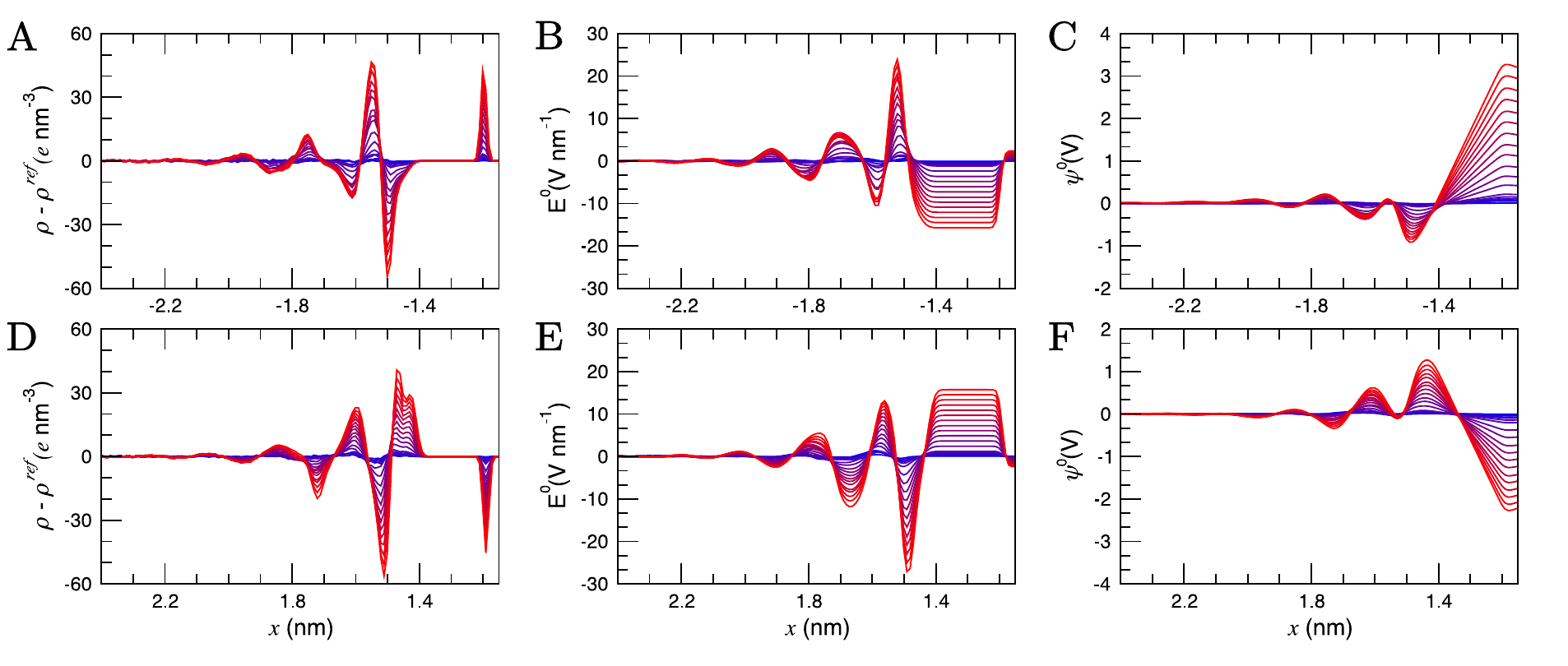}
  \caption[]{Excess charge density, $\rho^0=\rho(x)-\rho^{ref}(x)$ (A,D), electric field, E$^0$ (B,E), and electric potential, $\psi^0$ (C,F), in the electric double layer calculated from C$\mu$MD simulations of NaCl(aq) solutions in contact with positively (top) and negatively (bottom) charged graphite surfaces when the charge density of the double layer in the absence of a surface charge ($\rho^{ref}$) was first removed.  Positive and negative peaks around -1.2 and 1.2~nm in A and D, respectively, indicate the position of the graphite surface. Increasing surface charge density (in the range 0.004--1 $e$~nm$^{-2}$) is indicated by the
  blue $\rightarrow{}$red colour scale (see the key in Figure \ref{fgr:electrode-atomdensities} for a list of the sampled charge densities).} 
  \label{fgr:electrode-residual-electrics}
\end{figure}

\begin{figure}[ht]
\centering
  \includegraphics[width=0.75\linewidth]{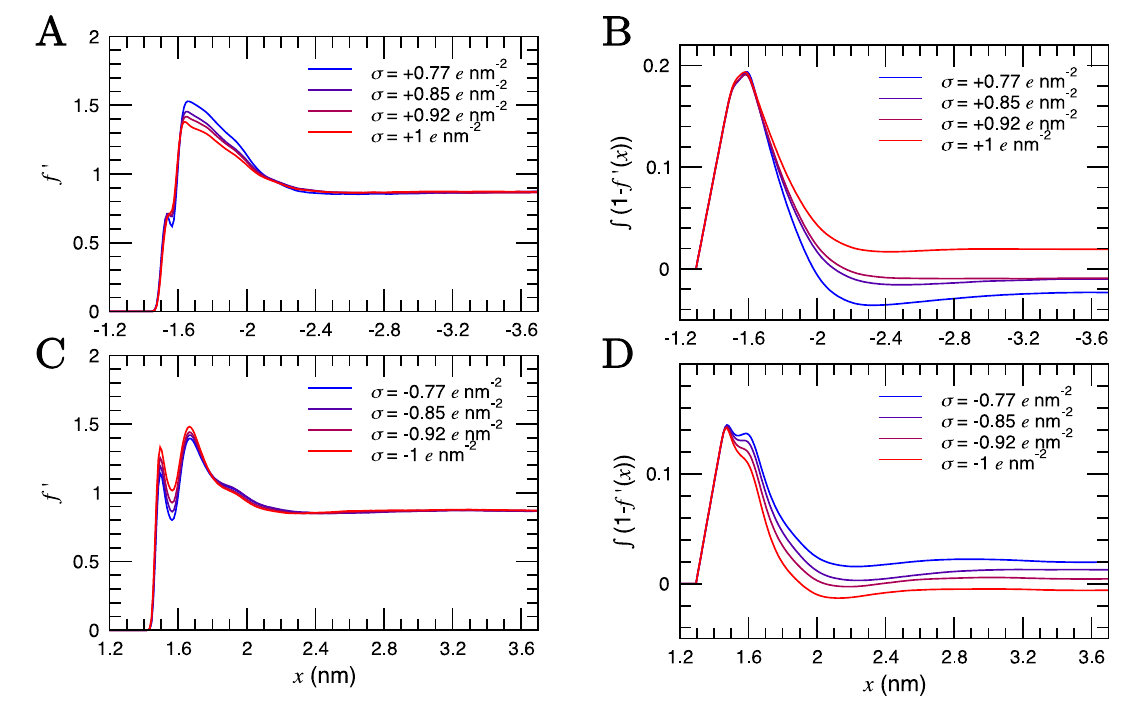}
  \caption[]{A and C: Electrode screening factor, $f'$, at a positively (A) and negatively (C) charged graphite surface with surface charge density, $\sigma$, indicated by the key. B and D: The cumulative integrals of $1-f'(x)$ using the data from A and C, respectively. The $x$ scale in A and B (i.e. at the positively charged surface) has been reversed to ease comparison with the profiles at the negatively charged surface.
  Note that in A and C, $f'$ consistently converge to a value of 0.87, i.e., lower than the predicted value of one. This is probably due to the influence of the long range contribution of the electrostatic interactions between charge carriers in solution and those on the opposite side of the graphite slab. These curves were, therefore, shifted so that $f'=1$ at large values of $x$ before calculating the curves in panels C and D.} 
  \label{fgr:electrode-screening-curves}
\end{figure}

\begin{figure}[ht]
\centering
  \includegraphics[width=0.6\linewidth]{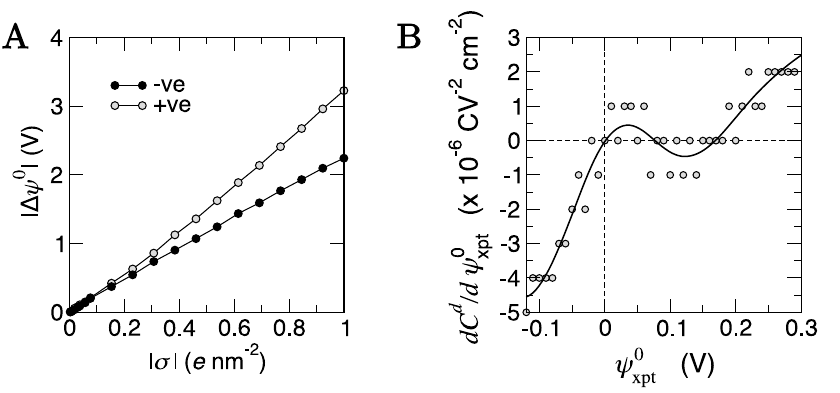}
  \caption[]{A: The potential difference across the double layer, $\Delta \psi^0$, when increasingly positive and negative surface charge densities, $\sigma$, were applied to the graphite surface in C$\mu$MD simulations.
  B: The derivative of the experimental differential capacitance, $C^d$, with respect to the applied potential, $\psi_{\mathrm{xpt}}^0$, calculated using the data in Figure \ref{fgr:mean_Cd} for the case of 1~M. The black line is a polynomial fit to the data which highlights the trends.} 
  \label{fgr:cap-deriv}
\end{figure}

\begin{figure}[ht]
\centering
  \includegraphics[width=0.325\linewidth]{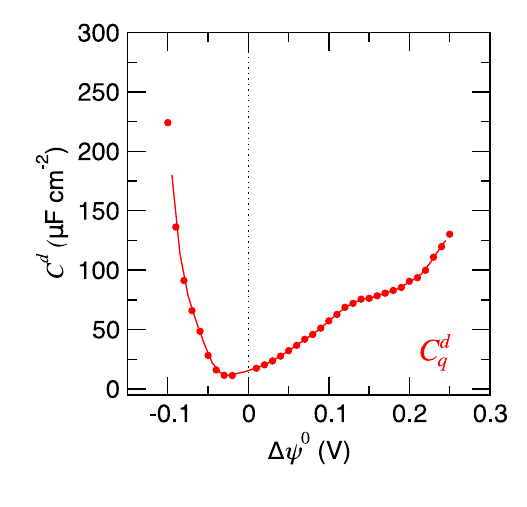}
  \caption[]{The `quantum' differential capacitance ($C^d_q$; due to the electronic response of graphite to charging) as a function of the potential difference relative to the potential of zero charge ($\Delta \psi^0$). $C^d_q$ was evaluated from the series $1/C^d_q = 1/C^d - 1/C^d_{dl}$ using the moving averages of simulation data ($C^d_{dl}$) and the experimental measurements ($C^d$) presented in the inset of Figure \ref{fgr:electrode-main}~F in the main paper.} 
  \label{fgr:cap-comp}
\end{figure}

\clearpage
\addcontentsline{toc}{section}{References}
\bibliographystyleSI{achemso}
\bibliographySI{graphite-nacl}

\end{document}